\documentclass[11pt]{article}
\usepackage{cospar}
\usepackage{url}

\usepackage{graphicx}
\usepackage[figuresright]{rotating}

\title{ MIGRATION OF ASTEROIDAL DUST}

\author{S. I. Ipatov\address{(1) NRC/NAS senior research associate, NASA/GSFC, Mail Code 685, 
Greenbelt, MD 20771, USA (current address); 
(2) Institute of Applied Mathematics, 
Miusskaya sq. 4, Moscow 125047, Russia}, 
J. C. Mather\address{NASA/GSFC, Mail Code 685, Greenbelt,
MD 20771, USA},
P. A. Taylor
\address{University of Maryland, College Park, MD 20742, USA}
}

\begin{document}

\maketitle

\begin{abstract}

We numerically investigate the migration of dust particles with initial orbits close to those of 
the numbered asteroids. The fraction
of silicate particles that collided 
with the Earth during their lifetimes varied from 0.2\% for 40 micron particles 
to 0.008\% for 1 micron particles. Almost all particles with diameter $d$$\ge$4 
microns collided with the Sun. 
The peaks in the migrating asteroidal dust particles' semi-major axis distribution
at the n/(n+1) resonances with Earth and Venus 
and the gaps associated with
the 1:1 resonances with these planets
are more pronounced for larger particles.

\end{abstract}

\section*{INTRODUCTION}

The orbital evolution of interplanetary dust particles has been simulated by several authors. 
Liou et al. (1996), Liou and Zook (1999), Gorkavyi et al. (2000), 
Ozernoy (2000), Moro-Martin and Malhotra (2002) considered the
migration of dust particles from the trans-Neptunian belt, Liou et al. (1999) from 
Halley-type comets, Liou et al. (1995) from Comet Encke, and
Reach et al. (1997), 
Kortenkamp and Dermott (1998) and Grogan et al. (2001) from asteroid families and 
short-period comets. Further references are presented in the above papers.
The main contribution to the zodiacal light is from particles
that range from 20 to 200 $\mu$m in diameter.
It has been estimated that 20,000 to 40,000 tons of small dust particles 
(micrometer-to-millimeter-sized) fall to the Earth every year with the mass distribution 
of dust particles 
peaking at about 200 $\mu$m in diameter.
Liou et al. (1995) showed that the observed shape of the zodiacal cloud
can be accounted for by a combination of about 1/4 to 1/3 asteroidal dust and
about 3/4 to 2/3 cometary dust (they considered production of dust from 
Encke-type parent bodies in the case when initial mean longitudes were random).

Analysis of the Pioneer 10 and 11 meteoroid detector data showed (Humes, 1980; Gr$\ddot u$n, 1994) 
that a population of $10^{-9}$ and $10^{-8}$ g ($\sim$10 $\mu$m)
particles has a constant spatial density 
between 3 and 18 AU. This result was confirmed in the runs by Liou and Zook (1999) 
for 23 $\mu$m trans-Neptunian dust particles. In the runs by 
Moro-Martin and Malhotra (2002) [Fig. 13], the surface (not spatial) 
density was approximately constant in this region. 
Spatial density of 1.4-10 $\mu$m particles obtained 
basing on the Voyager 1 data was
constant at 30 to 51 AU.  

Liou et al. (1996) noted that interstellar dust particles with an average size of 1.2 $\mu$m
can destroy dust particles formed in the solar system
and that the collisional lifetimes for 1, 2, 4, 9, 23 $\mu$m particles 
are 104, 49, 19, 4.8, 0.86 Myr, respectively.
In these size ranges mutual collisions are not as important as collisions
with interstellar grains.
Moro-Martin and Malhotra (2002) concluded that collisional destruction
is only important for kuiperoidal grains larger than about 6 $\mu$m 
(9 $\mu$m in Liou et al., 1996) 
and particles larger than 50 $\mu$m may also survive collisions because interstellar grains are too
small to destroy these in a single impact. These authors noted that for  silicate
particles of 1-40 $\mu$m in diameter the sublimation temperature ($\sim$1500 K) is
reached at $R$$<$0.5 AU, but for water ice particles the sublimation temperature 
($\sim$100 K) is reached at 27, 19, 14, 10, and 4.3 AU for the sizes of 3, 6, 11, 23, 
and 120 $\mu$m,respectively.

In the present paper we consider different initial orbits of dust particles, and
for the first time, investigate the collisional probabilities of migrating particles with
the planets based on a set of orbital elements during evolution. We also present
plots of the orbital elements of the migrating particles.

\section*{MODELS}

    Using the Bulirsh--Stoer method of integration, we investigated
the migration of dust particles under the gravitational influence of 
all of the planets (excluding Pluto), radiation pressure, Poynting--Robertson drag and solar wind 
drag, for the values of
the ratio between the radiation pressure force and the gravitational force
$\beta$ equal to 0.01, 0.05, 0.1, 0.25, and 0.4. For silicate particles 
such values of $\beta$ correspond to diameters of about 40, 9, 4, 
1.6, and 1 microns, respectively. 
$\beta$
values of 0.01 and 0.05 correspond to particle masses of $10^{-7}$ g 
and $10^{-9}$ g, respectively.
Burns et al. (1979) obtained $\beta$=$0.573 Q_{pr}/(\rho s)$, where
$\rho$ is the particle's density in grams per cubic centimeter,
$s$ is its radius in micrometers, and $Q_{pr}$ is the radiation
pressure coefficient ($Q_{pr}$ is close to unity for particles
larger than 1 $\mu$m). For water ice particles with
$\rho$=1 g cm$^{-3}$ particle diameters are 120, 23, 11, 6, and 3 $\mu$m for 
$\beta$=0.01, 0.05, 0.1, 0.2, and 0.4, respectively (Burns et al., 1979).
As did Liou et al. (1999) and Moro-Martin and Malhotra (2002), we assume 
the ratio of solar wind drag 
to Poynting--Robertson drag to be 0.35.
The relative error per integration step was taken to be less than $10^{-8}$.

Initial positions and velocities of the particles 
were the same as those of the first numbered main-belt asteroids (JDT 2452500.5),
i.e., dust particles are assumed to leave the asteroids with zero
relative velocity. 
For each $\beta$$\ge$0.05 we considered $N$=500 particles ($N$=250 
for $\beta$=0.01). In each run we took $N$=250, because for $N$$\ge$500 
the computer time per calculation for one particle was several times greater 
than for $N$=250. These runs were made until all of the particles
collided with the Sun or reached 2000 AU from the Sun. 
The lifetimes of all considered asteroidal dust particles were less than 0.8 Myr, except for
one particle with a lifetime of 19 Myr.  
We also made similar runs for the model without planets in order to 
investigate the role of planets in migration of dust particles.

\section*{COLLISIONS WITH PLANETS AND THE SUN}

In our runs planets were 
considered as material points, but, based on orbital elements 
obtained with a step $d_t$ of $\le$20 yr ($d_t$=10 yr for $\beta$ equal to 0.1 
and 0.25, and $d_t$=20 yr for other values of $\beta$), 
as in Ipatov and Mather (2003), we calculated the mean probability 
$P$=$P_{\Sigma}/N$ ($P_\Sigma$ is the probability for all $N$ 
considered particles) of a collision of a particle with a planet 
during the lifetime of the particle. We define
$T$=$T_{\Sigma}/N$ as the mean time during which the perihelion 
distance $q$ of a particle was less than the semi-major axis of the 
planet. Below, $P_{Sun}$ is the ratio of the number of particles that
collided with the Sun to the total number of considered particles,
and $P_{Sun}^{250}$ is the same ratio for the first 250 particles. 
$T_S^{min}$ and $T_S^{max}$ are the minimum and maximum values of the time until
the  collision of a particle with the Sun, and $T_{2000}^{min}$
and $T_{2000}^{max}$ are the minimum and maximum values of the time 
 when the distance between a particle and the Sun reached 2000 AU.
The values of $P_{Sun}^{250}$, $P_r$=$10^6 P$, $T$, $T_J$,
$T_S^{min}$, $T_S^{max}$, $T_{2000}^{min}$,
and $T_{2000}^{max}$  are shown in 
Table 1 for $N$=250 (for $\beta$=0.1 we present two runs with 250
different particles), and $P_{Sun}$ was obtained for all considered 
particles at a fixed $\beta$. 

The minimum time $T_S^{min}$ needed to reach the Sun is smaller for smaller particles, but 
the ratio  $T_S^{max}/T_S^{min}$ is much
greater 
and the ratio $T_{2000}^{max}/T_S^{max}$ is smaller 
for $\beta$$\ge$0.25 than for $\beta$$\le$0.1. 
For $\beta$=0.05, 498 particles collided with the Sun in less than 0.089 Myr, 
but two particles (with initial orbits close to those of the asteroids
361 and 499), which reached 2000 AU, lived for 0.21 Myr and 19.055 Myr, 
respectively. The latter object had perihelion near Saturn's orbit for a long time.


\begin{table}
\caption{Values of $T$, $T_J$, $T_S^{min}$, $T_S^{max}$, $T_{2000}^{min}$,
and $T_{2000}^{max}$ (in Kyr), $P_r$, $P_{Sun}^{250}$, and $P_{Sun}$ 
obtained for asteroidal dust particles for several values of $\beta$
(Venus=V, Earth=E, Mars=M)}

$ \begin{array}{lcccccccccccccc} 

\hline	

  & & & $V$ & $V$ & $E$ & $E$ & $M$ & $M$ & && &&& \\


\beta & P_{Sun}&P_{Sun}^{250} & P_r & T & P_r & T &  P_r & T & T_J & T_S^{min}&
T_S^{max}&T_S^{max}/T_S^{min}&T_{2000}^{min}&T_{2000}^{max} \\

\hline

0.01 & 1.000&1.000 & 1534 & 19.2 & 1746 & 44.2 & 127 & 99.9 & 0 &142& 422& 3.0&& \\
0.05 & 0.996&1.000 & 195  & 4.0  & 190  & 8.1 & 36.7 & 20.5 & 0 &30&  89&3.0&& \\
0.1  & 0.990&0.988 & 141  & 2.4  & 132  & 4.8 & 16.4 & 12.0 & 2.21 &16&44&2.8&138& 793 \\
0.1* & 0.990&0.992 & 366  & 2.4  & 279  & 4.8 & 20.9 & 12.0 & 0.92 &7.2&43&6.0&9& 534 \\
0.25 & 0.618&0.660 & 79.2 & 1.4  & 63.8 & 2.9 & 5.60 & 5.9  & 31.7 & 5.9&385&65&1.6&567 \\
0.4  & 0.316&0.324 & 12.4 & 1.5  & 8.0  & 2.5 & 0.72 & 8.8  & 32.3 &4.3&172&40&1.7& 288 \\

\hline
\end{array} $ 
\end{table}

For smaller particles (i.e., those with larger $\beta$), $P_{Sun}$ is smaller
and the probability
of collisions of particles with the 
terrestrial planets is smaller. The probability of a collision of a migrating dust particle 
with the Earth for $\beta$=0.01 is greater by a factor of 220 than for  $\beta$=0.4. 
These probabilities of collisions are in accordance with cratering records in 
lunar material and on the panels of the Long Duration Exposure Facility, 
which showed that the mass distribution of dust particles encountering Earth peaks at $d$=200 $\mu$m.

\section*{ORBITAL EVOLUTION}

Several plots of the distribution of 
migrating asteroidal particles in their orbital elements and  the distribution 
of particles with their distance $R$ from the Sun and their height $h$ above the initial 
plane of the Earth's orbit are presented in Figs. 1-8.  For Fig. 1-2
the number of  bins in the semi-major axis $a$ is 1000. For the other figures
the number of bins in $a$ or $R$ is 100, and the number
of bins in $e$ or $h$ is usually slightly less than 100. 
The width of a bin in $i$ is 1$^\circ$. 

In Figs. 1-2, 4-8 for calculations of 
orbital elements we 
added the coefficient (1-$\beta$) to the mass of the Sun,
which is due to the radiation pressure. 
If  not stated otherwise, we consider such orbital elements.
Fig. 3 is similar to Fig. 4, but in Fig. 3 
(and also in the figures presented by Ipatov et al., 2003)
for calculations of osculating orbital elements we did not 
add this coefficient 
to the mass of the Sun,
(i.e., to transfer from rectangular coordinates to
orbital elements we used the same formulas as those for massive bodies).
For transformations between semi-major axes and eccentricities
for the above two systems of orbital elements we used the
formulas obtained by Kortenkamp and Dermott (1998).
In total, Figs. 3 and 4 are similar, but they differ in some details.
For example, for $\beta$=0.4 in Fig. 4 there are some pairs of $a$ and $e$
corresponding to the perihelion near Saturn's orbit. There are no
such pairs in Fig. 3 for $\beta$=0.4.

\begin{figure}
\includegraphics[width=90mm]{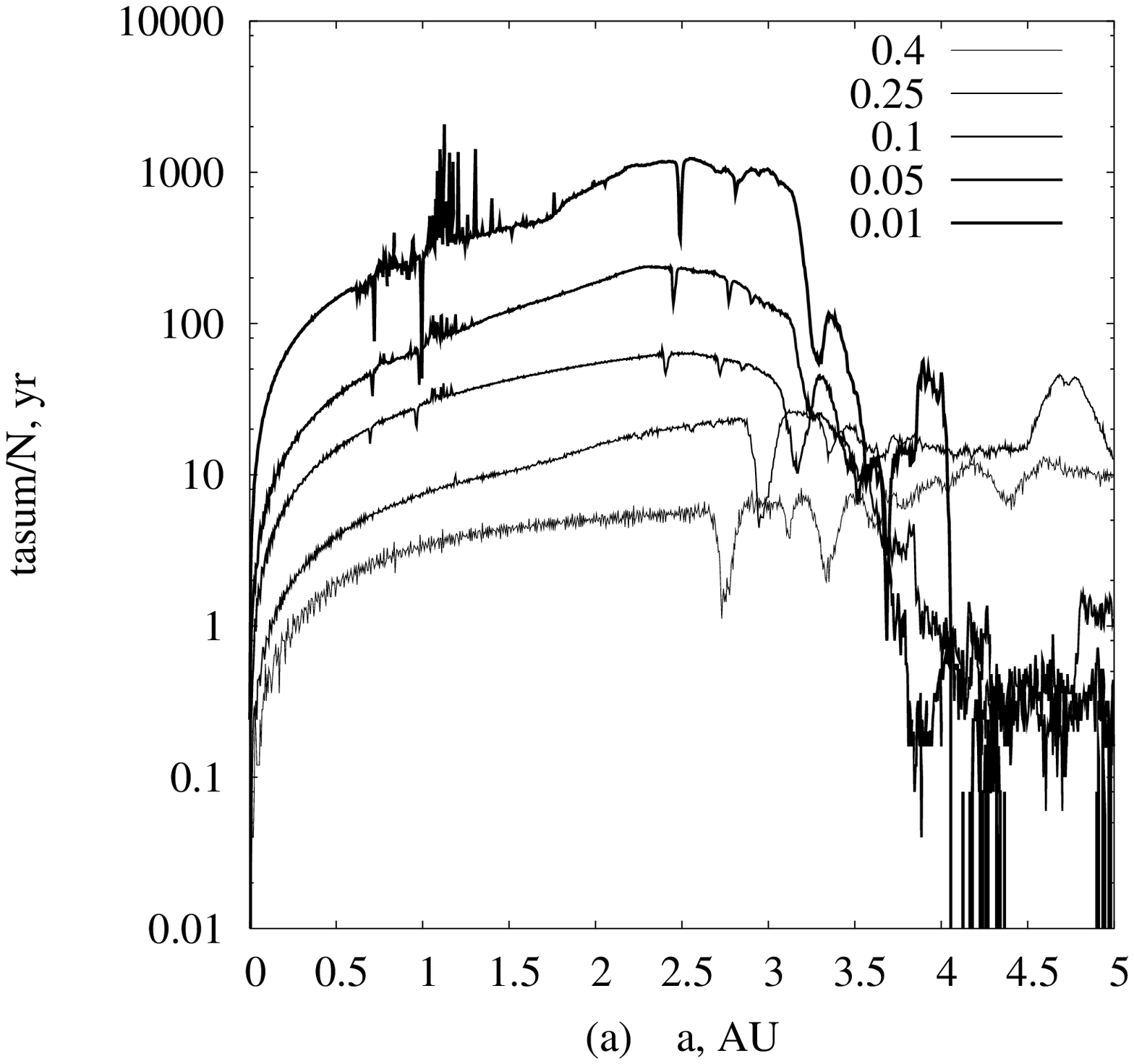}
\includegraphics[width=90mm]{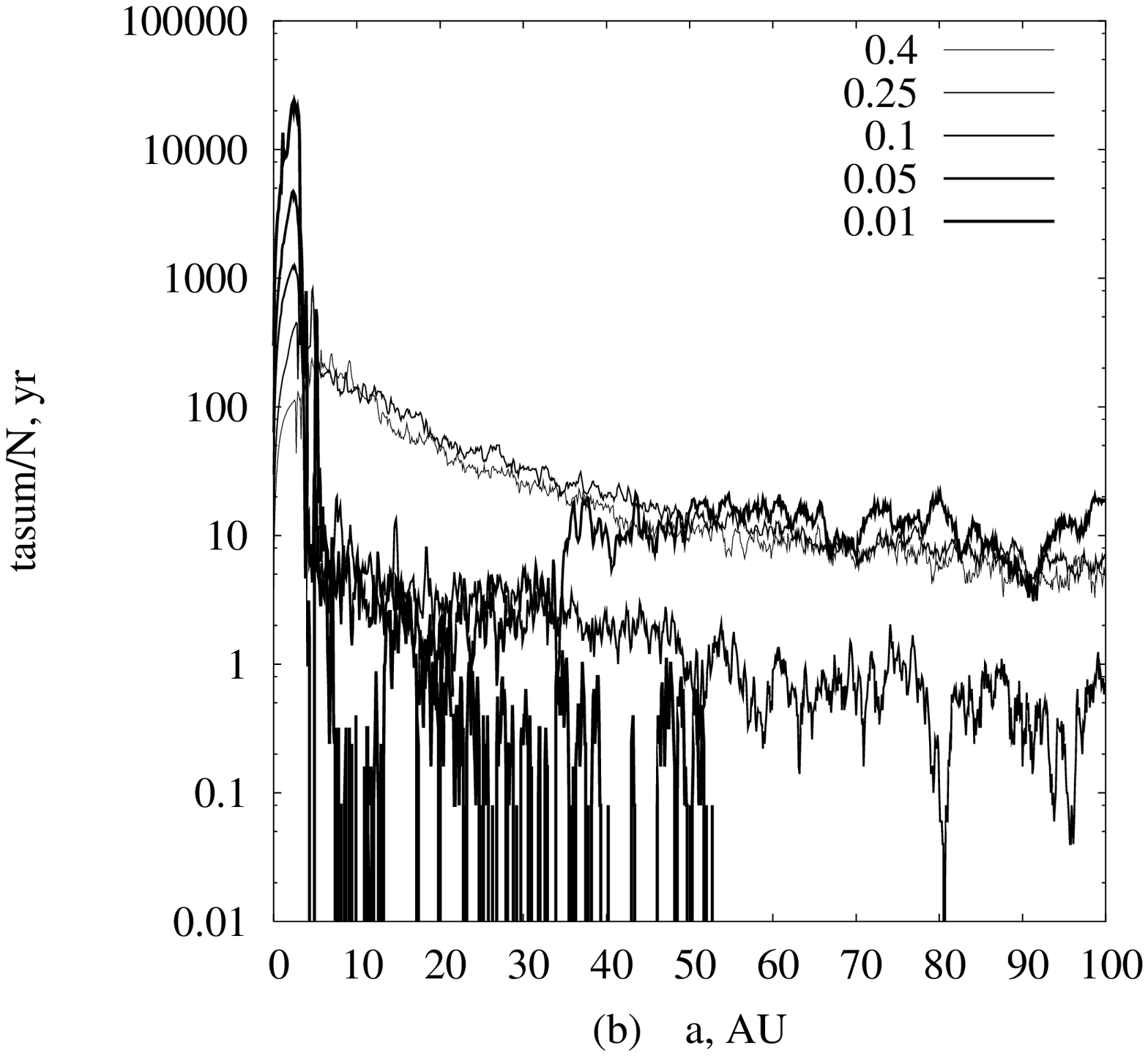}
\includegraphics[width=180mm]{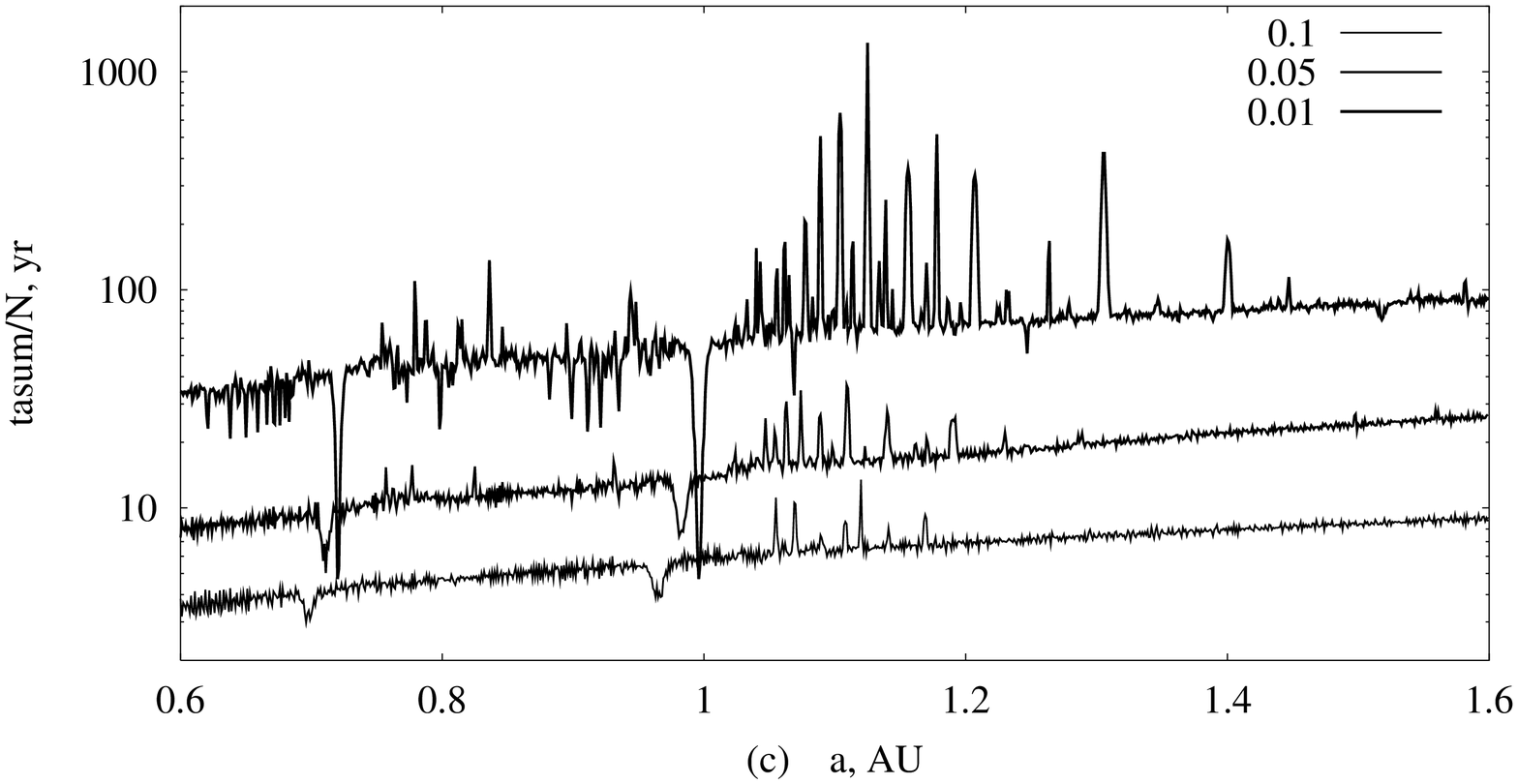}

\caption{
Mean time $t_a$ (the total time {\it tasum} divided by the number 
$N$ of particles) during which an asteroidal dust particle 
had a semi-major axis in an interval with a width of (a) 0.005 AU, (b) 0.1 AU, or (c) 0.001 AU. 
The values of $t_a$ at 2 AU are greater for 
smaller $\beta$. Curves plotted at 40 AU are (top-to-bottom) 
for $\beta$ equal to 0.25, 0.4, 0.05, 0.1, and 0.01. For 
calculations of orbital elements we added the coefficient 
(1-$\beta$) to the mass of the Sun, which is due to the radiation 
pressure (the same orbital elements are used
in Figs. 1-2, 4-8). Initial velocities and coordinates of dust particles 
were the same as those of the first $N$=500 numbered asteroids 
($N$=250 for $\beta$=0.01) at JDT 2452500.5.
}

\end{figure}%

\begin{figure}
\includegraphics[width=90mm]{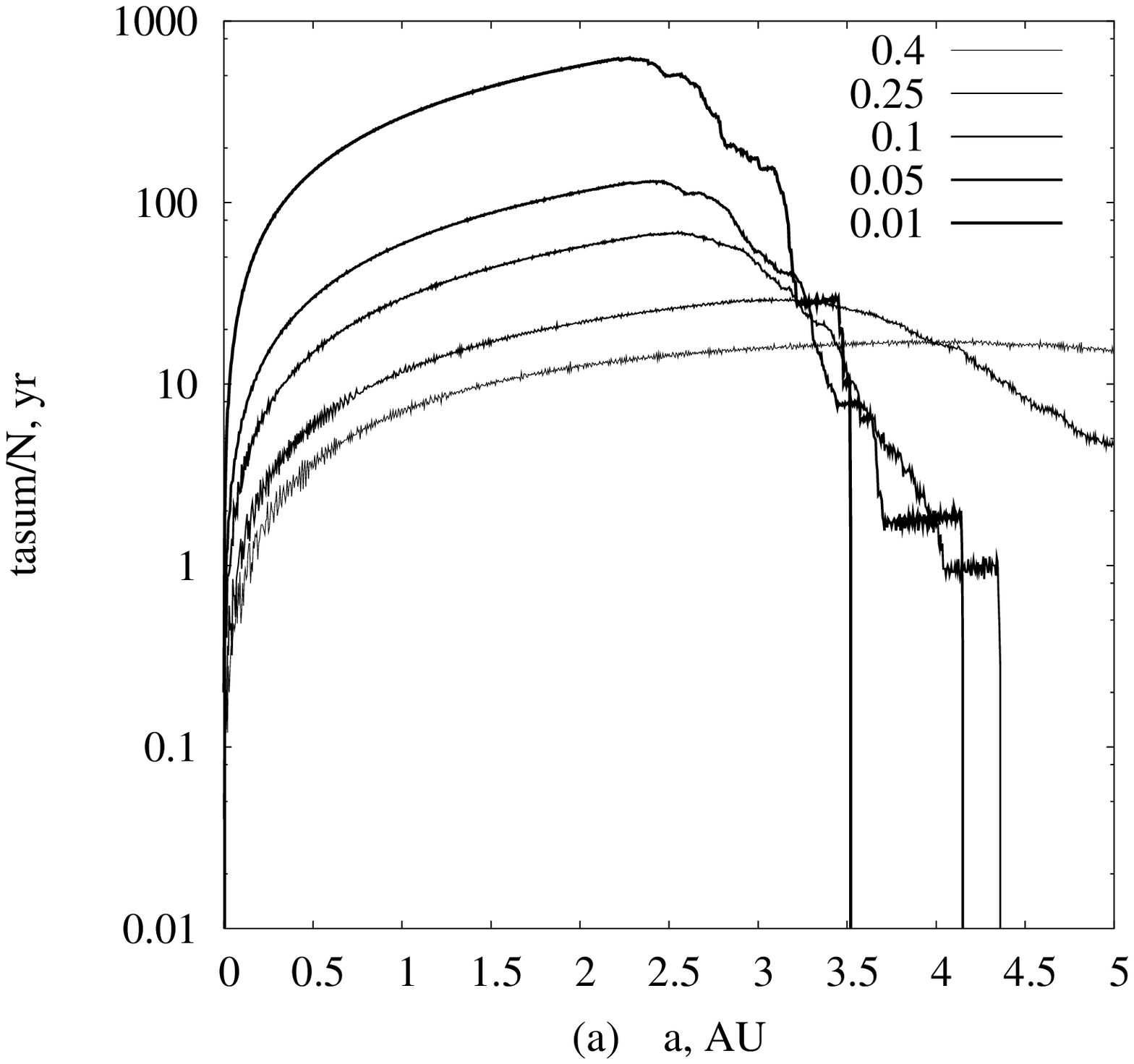}
\includegraphics[width=90mm]{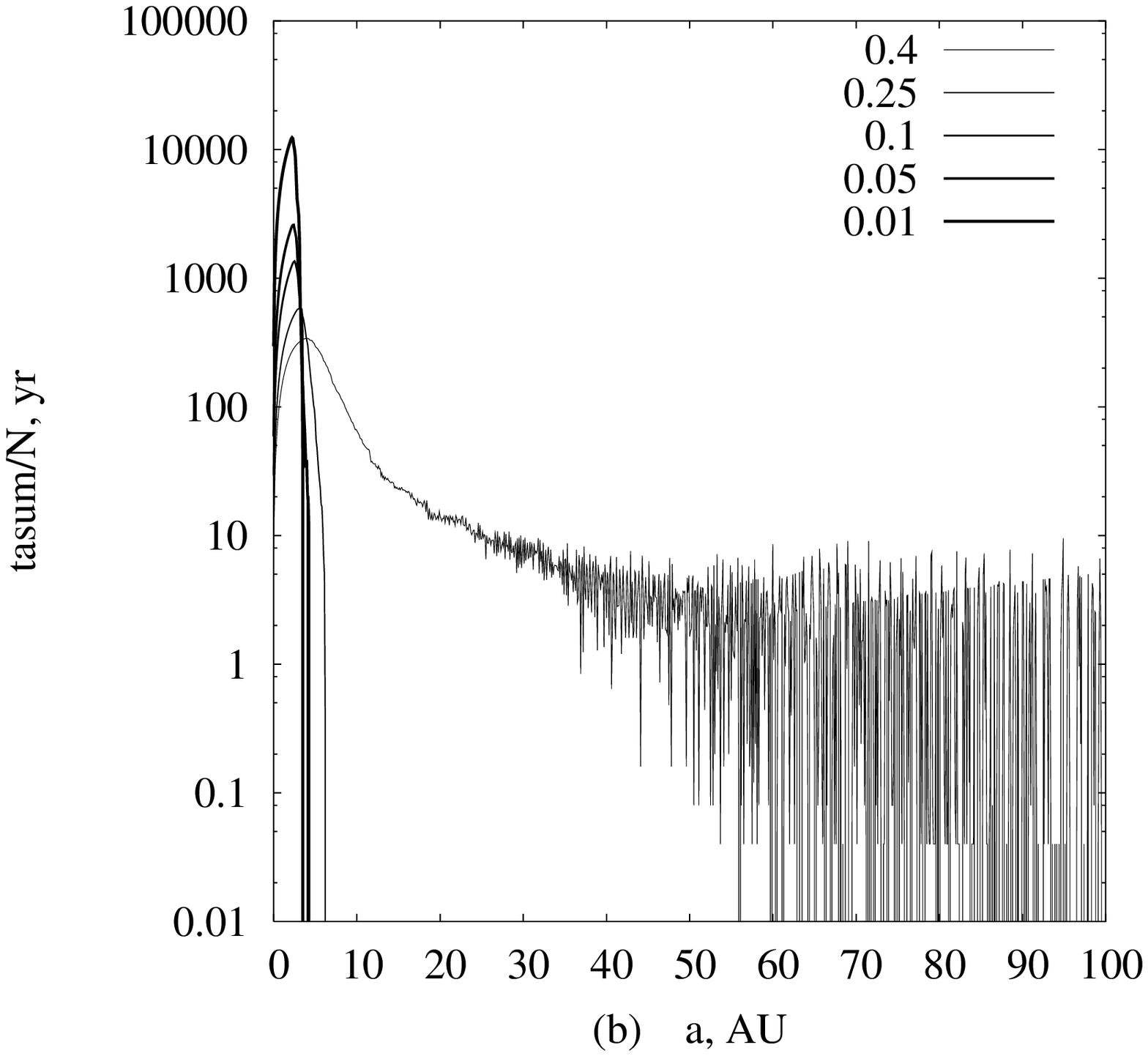}

\caption{
Same as for Fig. 1 but without gravitational influence of 
planets and for a smaller number of initial dust particles 
($N$=250 at $\beta$$\ge$0.05, and $N$=100 at $\beta$$\le$0.01). 
The values of $a$$>$6.2 AU were reached only at $\beta$=0.4. 
}

\end{figure}%

We obtained that the mean time $t_a$ 
(the total time divided by the number $N$ of particles) 
during which an asteroidal dust particle had 
a semi-major axis $a$ in an interval of fixed width is 
greater for smaller $\beta$ at semi-major axes $a$$<$3 AU (exclusive of the gap
at $a$=1 AU and $\beta$=0.01).
In Fig. 1b curves plotted at 40 AU are (top-to-bottom) 
for $\beta$ equal to 0.25, 0.4, 0.05, 0.1, and 0.01.
For $\beta$$\le$0.1 the values 
of $t_a$ are much smaller at $a$$>$3.5 AU than at $1$$<$$a$$<$3 AU, and they are usually a 
maximum at $a$$\approx$2.3 AU. For $\beta$=0.01 the local maxima of $t_a$ 
corresponding to the  6:7, 5:6, 3:4, and 2:3 resonances with the Earth
are greater than the maximum at 2.4 AU. 
There are several other local maxima (Fig. 1c) corresponding to the 
n/(n+1) resonances 
with Earth and Venus (e.g., the 7:8 and 4:5 resonances with Venus).
Dermott et al. (1994a-b) considered $\beta$=0.037 and 
showed that these resonances with the Earth
cause the Earth's asteroidal ring.
The greater the $\beta$, the smaller the local maxima
corresponding to these resonances.
At $\beta$$\le$0.1
there are gaps with $a$ a little smaller than the semi-major 
axes of Venus and Earth; the greater the $\beta$, the smaller the corresponding
values of $a$. 
A small gap for Mars is seen only at $\beta$=0.01.
There are also gaps corresponding to the 3:1, 5:2, and 2:1 resonances
with Jupiter.

For all considered $\beta$, $t_a$ 
decreases considerably  with a decrease of $a$ at $a$$<$1 AU and usually decreases with 
an increase of $a$ at $a$$>$5 AU (Fig. 1). 
Relatively large values of $t_a$ at $a$$>$40 AU for $\beta$=0.05
are due to only one particle.
For $a$$>$5 AU the values of $t_a$ are usually a little greater 
 at $\beta$=0.25 than those at $\beta$=0.4. The number
of particles, which got $a$$>$5 AU, at $\beta$=0.25 is smaller
than at $\beta$=0.4, but they move more slowly to 2000 AU than at 
$\beta$=0.4.
Analyzing Fig. 1a, we can conclude that the portion of
greater particles among all particles is on average greater at 
the zone of the terrestrial planets 
than in the asteroid belt.

In Fig. 2 we present the results obtained for the model without planets. In this case,
migration outside 5 AU is smaller than in the model with planets 
and, of course, there are no peaks
and gaps caused by planets (compare Figs. 1 and 2). 
For the model without planets, the values of $T_S^{min}$
were about the same as those presented in Table 1,
but the values of $T_S^{max}$ sometimes were smaller.
The values of $P_{Sun}^{250}$ were 0.908 and 0.548 for $\beta$=0.25
and  $\beta$=0.4, respectively (greater than for the model with planets).
For  $\beta$$\le$0.1 all particles collided with the Sun.

We now return to the model with planets.
At $a$$<$4 AU the maximum eccentricities for $\beta$$\ge$0.25
were greater than those for $\beta$$\le$0.1 (Figs. 4-5). 
At $\beta$=0.01 some particles got into the 1:1 resonance  with Jupiter.
For $a$$>$10 AU perihelia were usually near Jupiter's orbit (for $\beta$=0.05 and $\beta$=0.25 
also near Saturn's orbit). 
In almost all cases, the inclinations $i$$<$$50^o$; at $a$$>$10 AU the
maximum $i$ was smaller for smaller $\beta$ (Fig. 6).

In Fig. 5 we present the distributions of dust particles in $a$ and $e$, and in $a$ and $i$ for
$\beta$=0.1. The plots on the left were obtained for initial 
positions and eccentricities close to those
of the asteroids with numbers 1-250, and the plots on the right, 
to those of the asteroids with numbers 251-500 
(Figs. 1-4, 6-8 were obtained using all particles).
In total, the left and the right plots are similar, but for the left plots
particles spent more time outside 5 AU.
At $\beta$=0.05 none of the first 250 particles reached Jupiter's orbit, but
two particles with numbers 361 and 499 migrated to 2000 AU having their perihelia
near the orbits of Jupiter and Saturn (Fig. 4).

Usually there are no particles 
with $h/R$$>$0.7 at $R$$<$10 AU, with $h/R$$>$0.25 at $R$$>$20 AU for $\beta$$\le$0.1, and 
with $h/R$$>$0.5 at $R$$>$50 AU for $\beta$$\ge$0.25. 
For $\beta$$\ge$0.25 at 
$R$$<$1000 AU 
the entire region with $h/R$$<$0.3 was not empty (Fig. 7). 

The total time spent by 250 particles in inner-Earth, Aten, Apollo and Amor
orbits was 5.6, 1.4, 4.5, and 7.5 Myr at $\beta$=0.01, and
0.09, 0.08, 0.48, and 0.76 Myr at $\beta$=0.4, respectively. 
The spatial density of a dust cloud and its luminosity
(as seen from outside) were greater for smaller $R$.
For example, depending on $\beta$ they were by a factor of 2.5-8 and 7-25 
(4 and 12-13 at $\beta$$\le$0.05)
greater at 1 AU than at 3 AU for the spatial density and luminosity, respectively. 
This is in accordance with the observations for the inner solar system
(inversion of zodiacal light observations by the Helios spaceprobe in the 
inner solar system revealed a particle density $n(R)\propto R^{-1.3}$,
Pioneer 10 observations between the Earth's orbit and the asteroid belt yielded 
$n(R)\propto R^{-1.5}$ and IRAS observations have yielded $n(R)\propto R^{-1.1}$,
Reach, 1992; the intensity $I$ of zodiacal light falls off with heliocentric distance
$R$ as $I$$\sim$$R^{-\gamma}$, with $\gamma$=2 to 2.5; beyond about 3 AU
zodiacal light was no longer observable above the background light, Gr$\ddot u$n, 1994).
As in Liou and Zook (1999), we approximately defined the brightness of each 
particle  as $R^{-2}$.  Beyond Jupiter's orbit even the number of asteroidal 
particles at some distance $R$ from the Sun is smaller for greater $R$, so 
asteroidal dust particles cannot explain the constant space density of dust 
particles at $R$$\sim$3-18 AU. Besides  dust particles that came from the 
trans-Neptunian belt, many dust particles at such distances could have come 
from the comets that pass through this region.

\section*{MIGRATION OF TRANS-NEPTUNIAN AND COMETARY DUST PARTICLES}

We also began a series of runs in which 
initial positions and velocities of the particles 
were the same as those of the first trans-Neptunian objects (JDT 2452600.5),
and our initial data were different from those in previous papers.
The number of particles was $N$=250 at $\beta$=0.4, $N$=100 at 
$\beta$=0.2 and $\beta$=0.1, and $N$=50 at $\beta$=0.05. We store orbital elements
with a step of 100 yr. However,
these runs have not  been finished at the present time.

The distributions of dust particles with semi-major axis $a$ and 
eccenticity $e$ or inclination $i$, and 
with distance $R$ from the Sun and 
height $h$ above the initial plane of the Earth's orbit are presented in Fig. 8.
The left plots were obtained for $\beta$=0.05, $N$=50, for a
 time interval of 2 Myr, and the right plots were made for
$\beta$=0.1, $N$=100, and $t$$\le$3.5 Myr. For both values of $\beta$,
bodies with $e$$>$0.4 had their perihelia mainly near the semi-major axis of Neptune.
There were also bodies with perihelion distance
$q$ near the semi-major axis of Uranus. 
Inclinations were less than 35$^\circ$.
For $\beta$=0.1, there was an increase in $i$ at $a$$\approx$10-12 AU. 

The values of $h$ were maximum (26 AU) at $R$$\sim$47-50 AU for $\beta$=0.05,
and were maximum (30 AU) at $R$$\sim$57 AU for $\beta$=0.1.
For $\beta$=0.05 at $t$$\le$2 Myr and for $\beta$=0.1 at $t$$\le$3 Myr, 
the values of  the number $n_R$ of particles at distance $R$,
and also the values of $n_R/R$ (surface density), $n_R/R^2$ (spatial density), and
even $n_R/R^3$ (similar to luminosity) were maximum at $\sim$40-45 AU,
but at that time only a few objects had reached $R$$<$20 AU (Fig. 6).

At $\beta$=0.4  most particles were outside 50 AU after only 0.02 Myr.
If we consider positions of particles for $t$$\le$0.05 Myr, then 
84\% of them had $R$$>$100 AU. At that time 
$n_R/R$, $n_R/R^2$, and $n_R/R^3$ were maximum at $\sim$40-55 AU.
For $\beta$=0.2 at $t$$\le$1 Myr, the luminosity was maximum 
at 30-65 AU (differed by a factor of less than 3 in this region),
but $R$$>$100 AU for only 1\% of orbital positions.

For each $\beta$, the mean eccentricity $e_m$ of former trans-Neptunian dust particles
increased with $a$ for
$a$$>$48 AU, and it exceeded 0.5 at	 $a$$>$60 AU.
Bodies that migrated inside Neptune's orbit had $e_m$$<$0.2.
We usually obtained $h/R$$<$0.5, and 
in the zone of the Edgeworth--Kuiper belt
the ratio of the number of particles at $h$=$kR$ 
dropped usually to 10\% of the number at $h$=0  at $k$=0.1, but sometimes at $k$=1/4. 

The trans-Neptunian belt is considered to be the main source of Jupiter-family comets. 
As Jupiter-family comets produce much dust, they can produce trans-Neptunian dust just
inside Jupiter's orbit. Some of these comets can reach typical near-Earth objects' 
orbits (Ipatov and Mather, 2003). The total mass of comets inside Jupiter's orbit
is much smaller than the total mass of asteroids, but a comet produces more dust 
per unit minor body mass than an asteroid.

\section*{CONCLUSIONS}

We investigated collision probabilities of migrating asteroidal dust particles
with the terrestrial planets during the lifetimes of these particles.
These probabilities were considerably larger for larger particles, 
which is in accordance with the analysis of microcraters.
 Almost all particles with diameter $d$$\ge$4 $\mu$m
collided with the Sun. In almost all cases, inclinations $i$$<$50$^o$, and
at $a$$>$10 AU maximum $i$ was smaller for larger asteroidal particles.
The spatial density of asteroidal particles decreases considerably at $a$$>$3 AU,
so the portion of asteroidal particles among other particles beyond Jupiter's orbit is small. The peaks in distribution
of migrating asteroidal dust particles with semi-major axis
corresponding to the n/(n+1) resonances with Earth and Venus 
and the gaps associated with the 1:1 resonances with these planets
are more pronounced for larger particles.

\section*{NOTES}

This version of the paper differs  from the first version submitted
to the Proceedings
of the international conference "New trends in astrodynamics and applications" (20-22 January 2003, University of Maryland, 
College Park). We do not know which version will appear in the proceedings.

\section*{ACKNOWLEDGEMENTS}

     This work was supported by NRC (0158730), NASA (NAG5-10776), INTAS (00-240), 
and RFBR (01-02-17540).

\begin{figure}
\includegraphics[width=90mm]{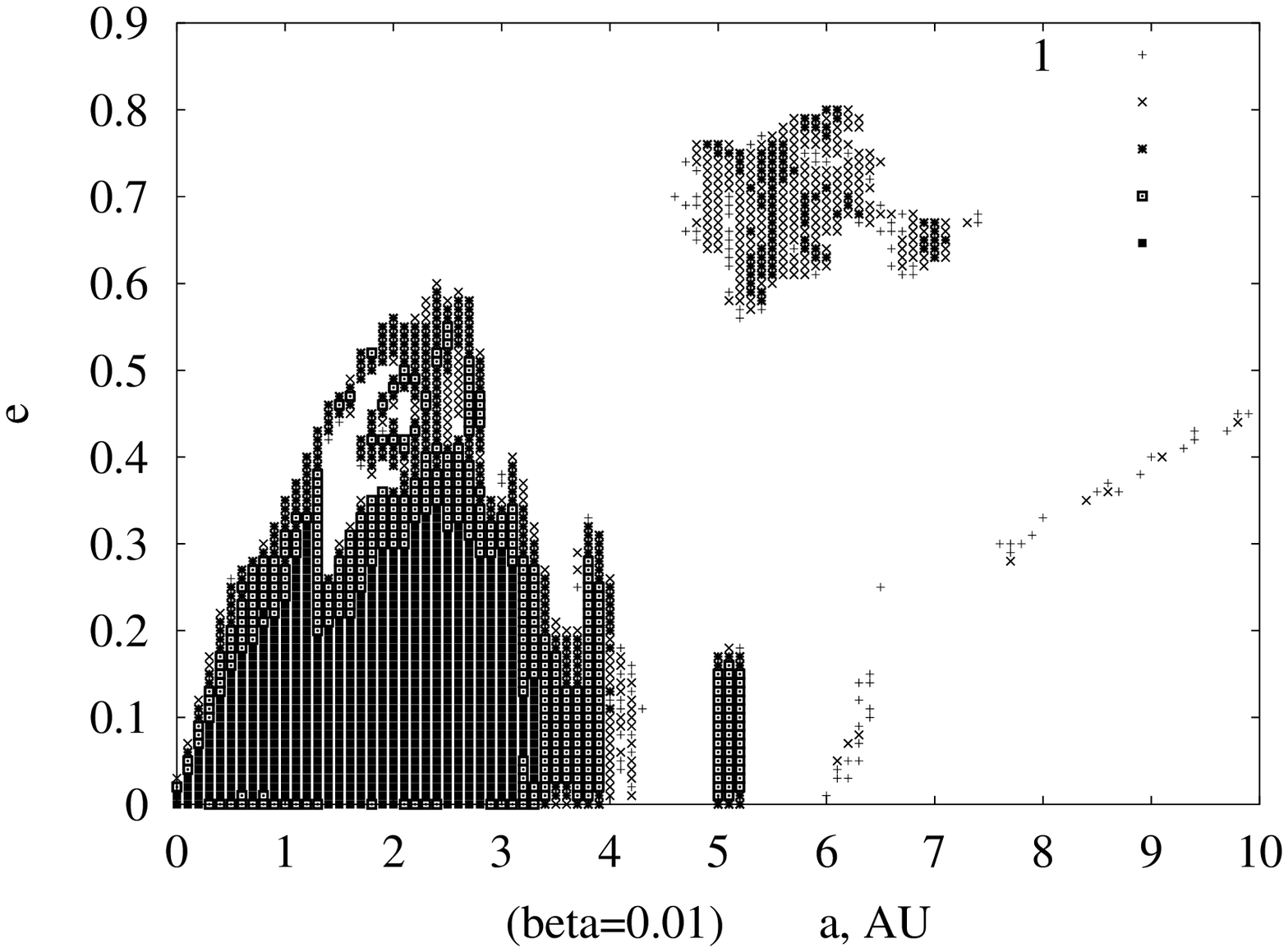}
\includegraphics[width=90mm]{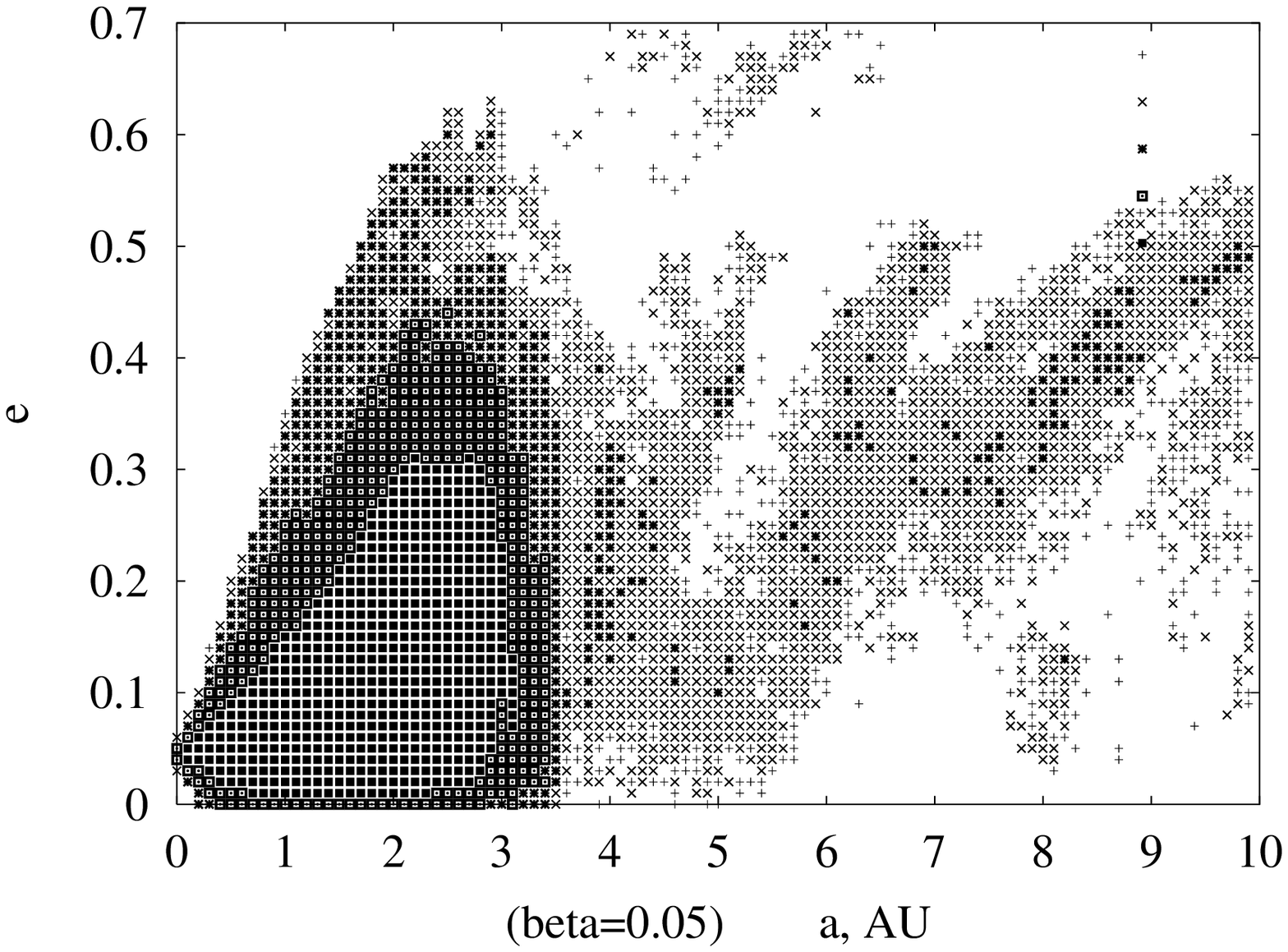}
\includegraphics[width=90mm]{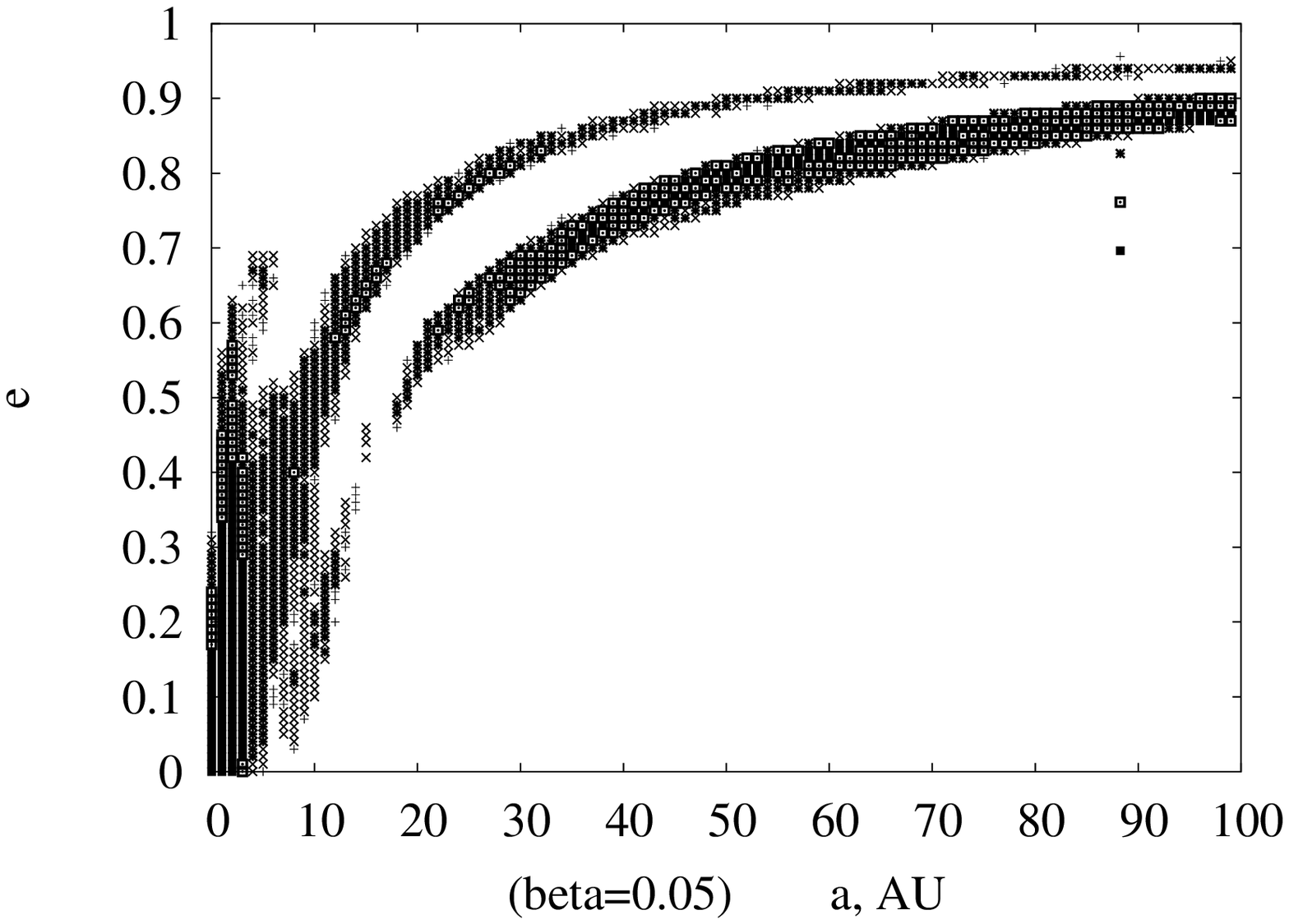}
\includegraphics[width=90mm]{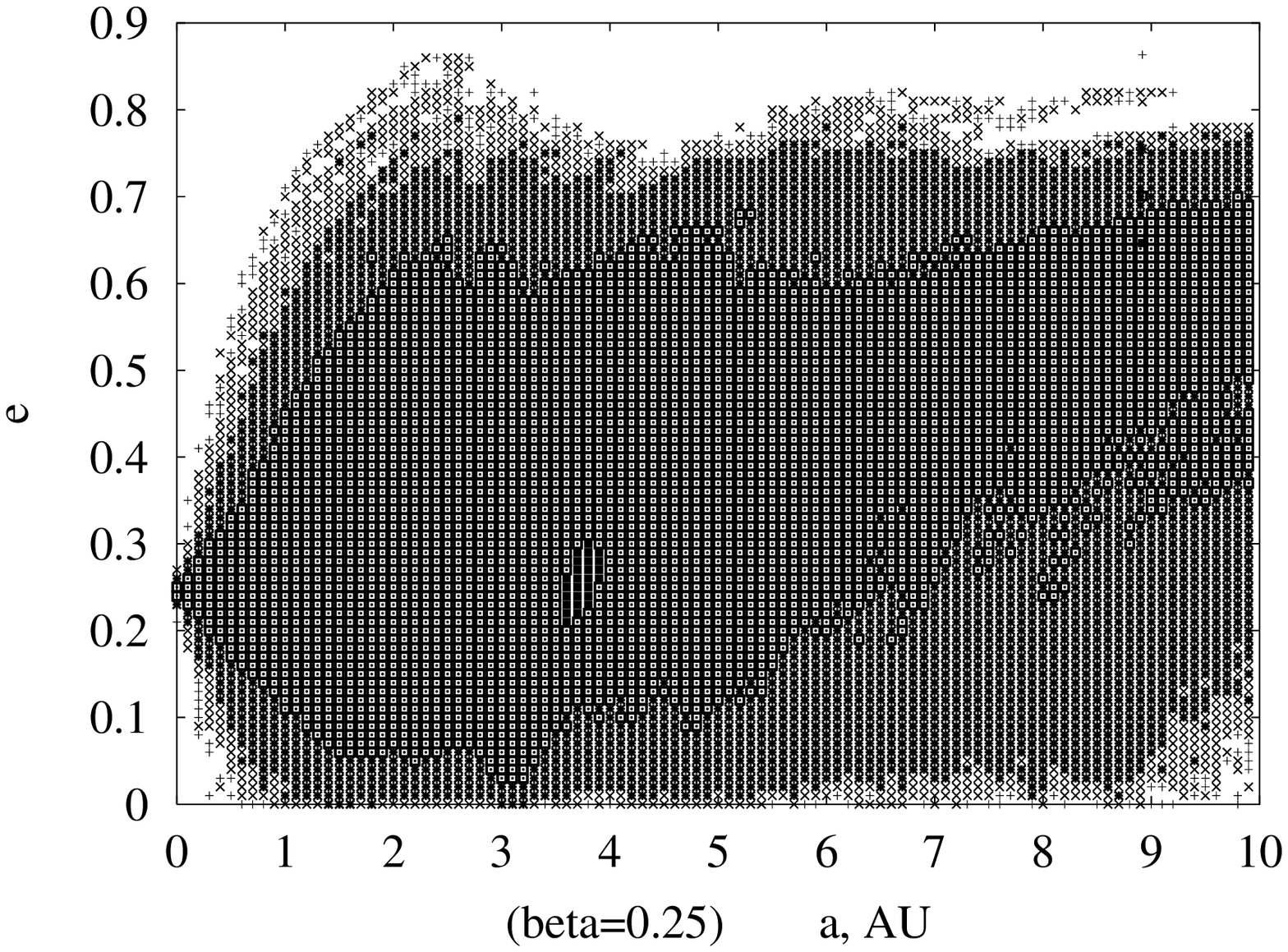}
\includegraphics[width=90mm]{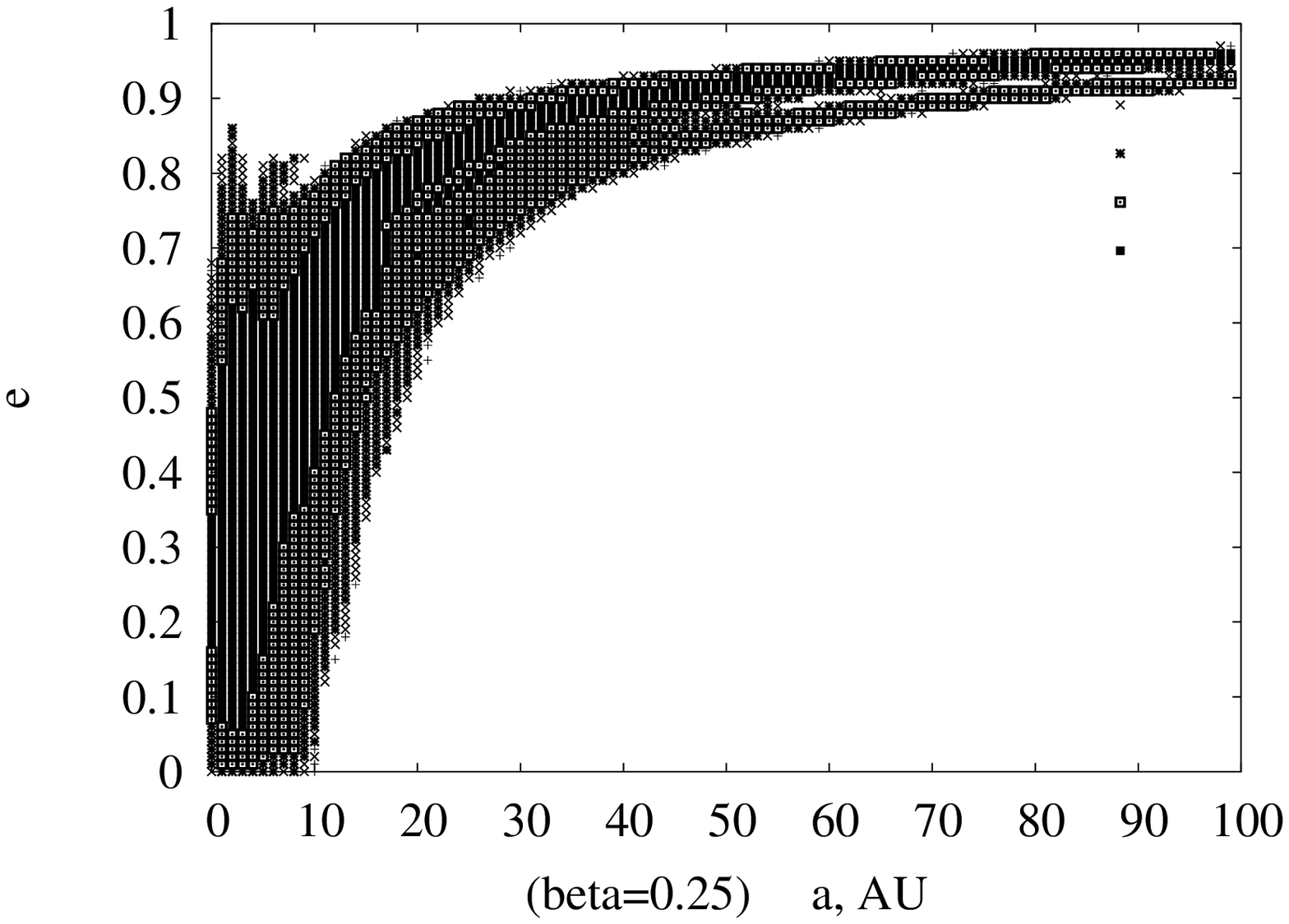}
\includegraphics[width=90mm]{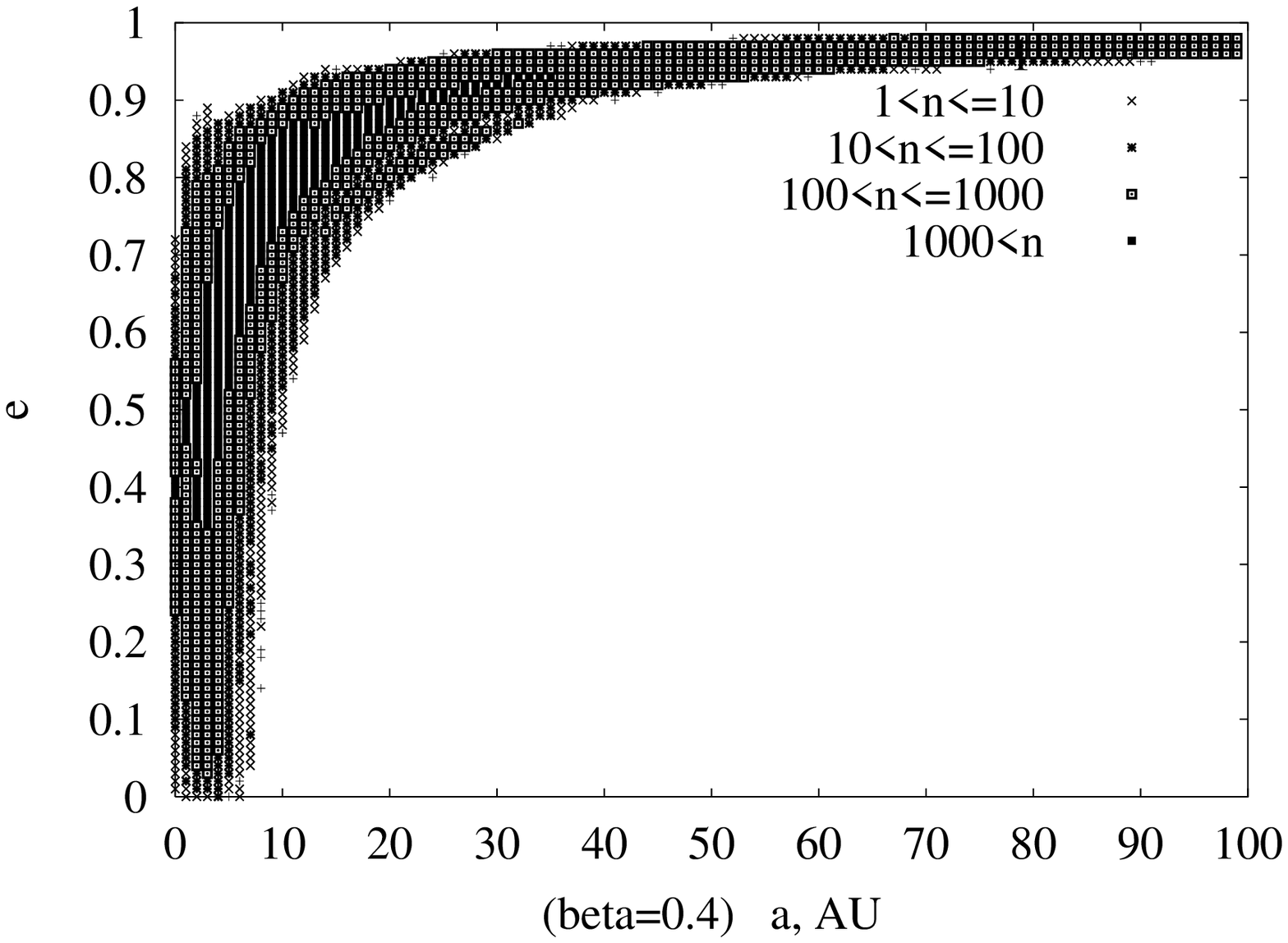}

\caption{Distribution of asteroidal dust particles with semi-major axis and eccentricity
(designations of the number of particles in one bin are 
the same in Figs. 3-7).
For the transfer from rectangular coordinates to
orbital elements we used the same formulas as those for massive bodies.
For Figs. 1-2, 4-8 for calculations of orbital elements we 
added the coefficient (1-$\beta$) to the mass of the Sun,
which is due to the radiation pressure. 
}

\end{figure}%

\begin{figure}
\includegraphics[width=90mm]{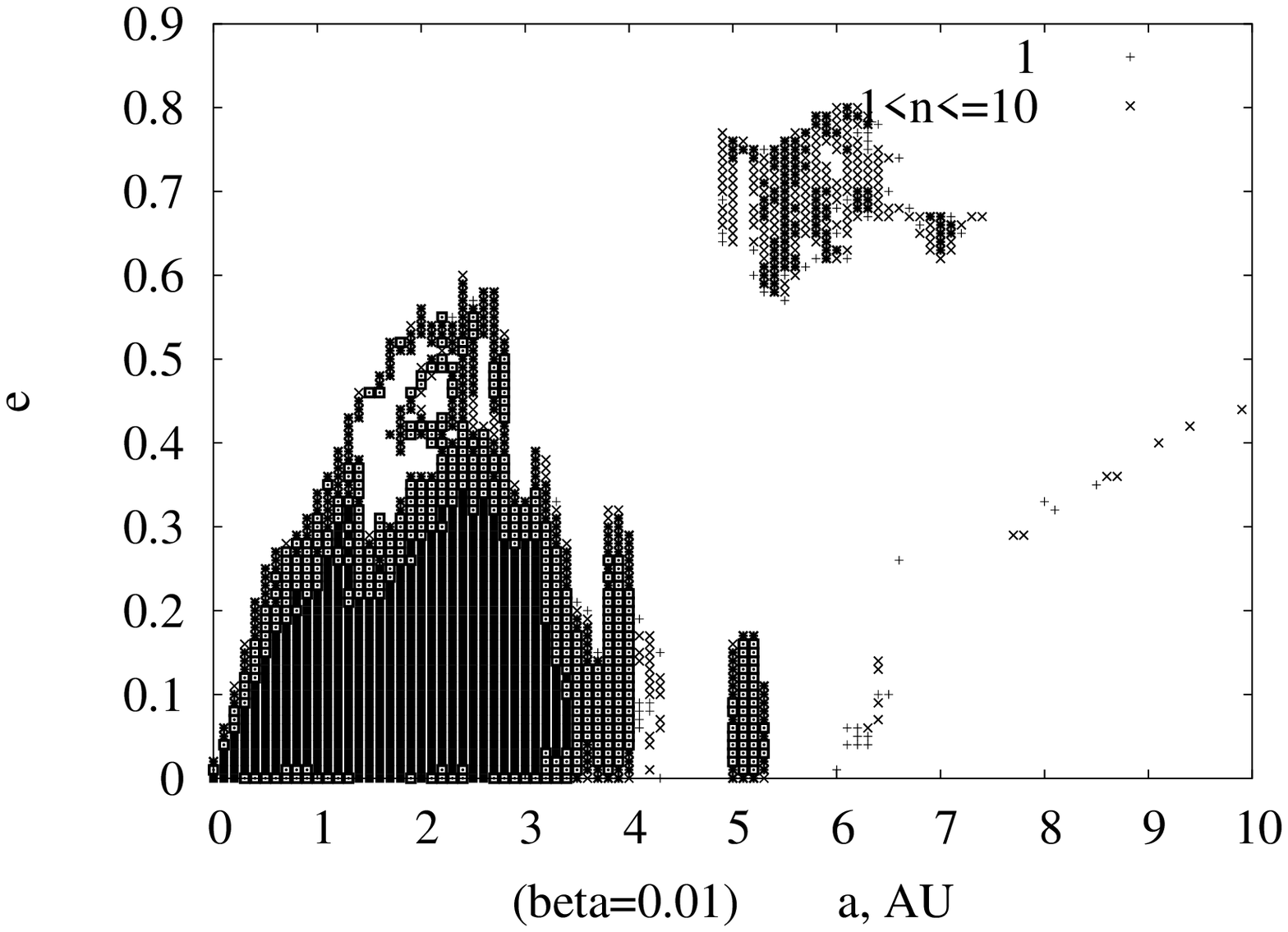}
\includegraphics[width=90mm]{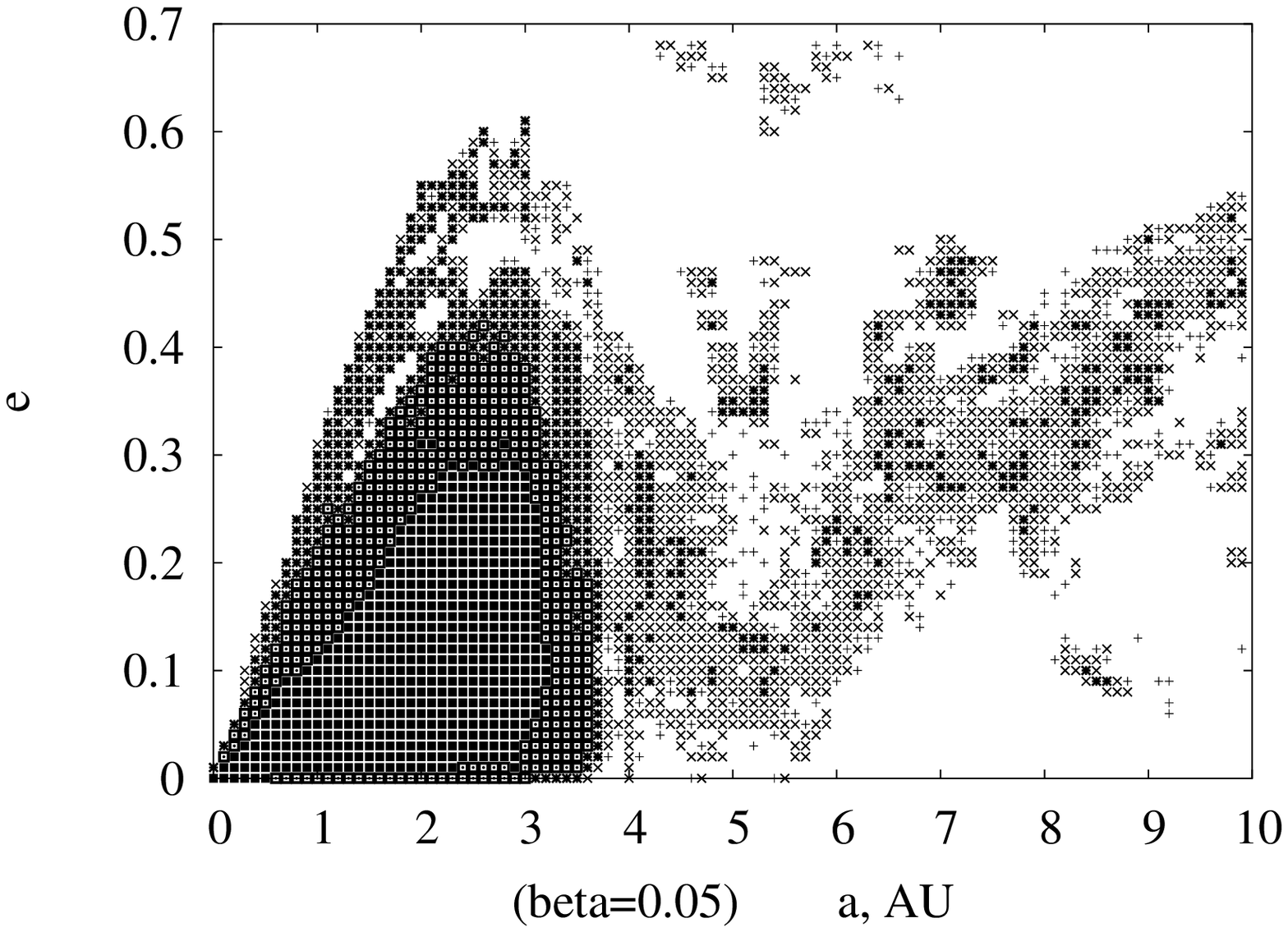}
\includegraphics[width=90mm]{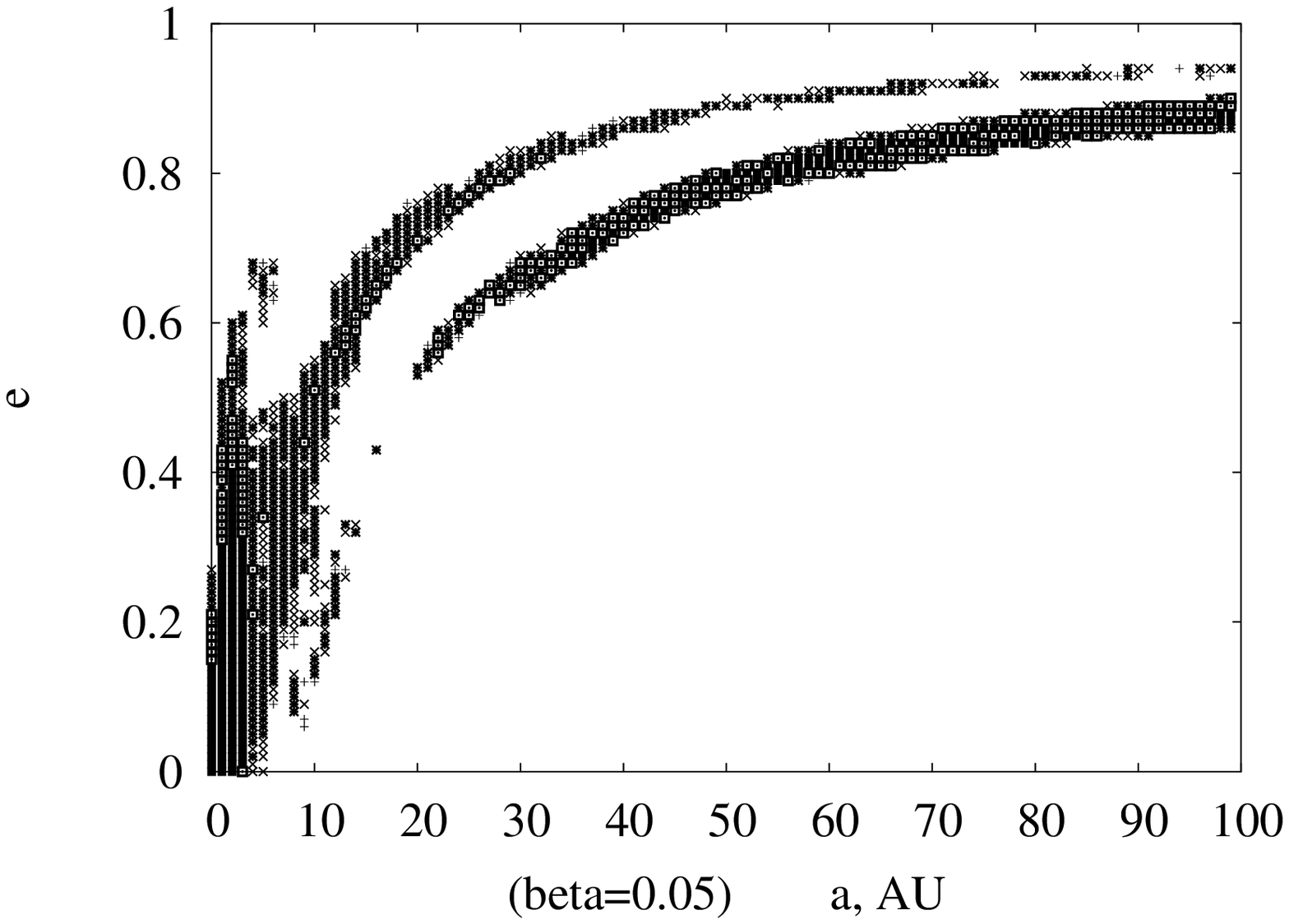}
\includegraphics[width=90mm]{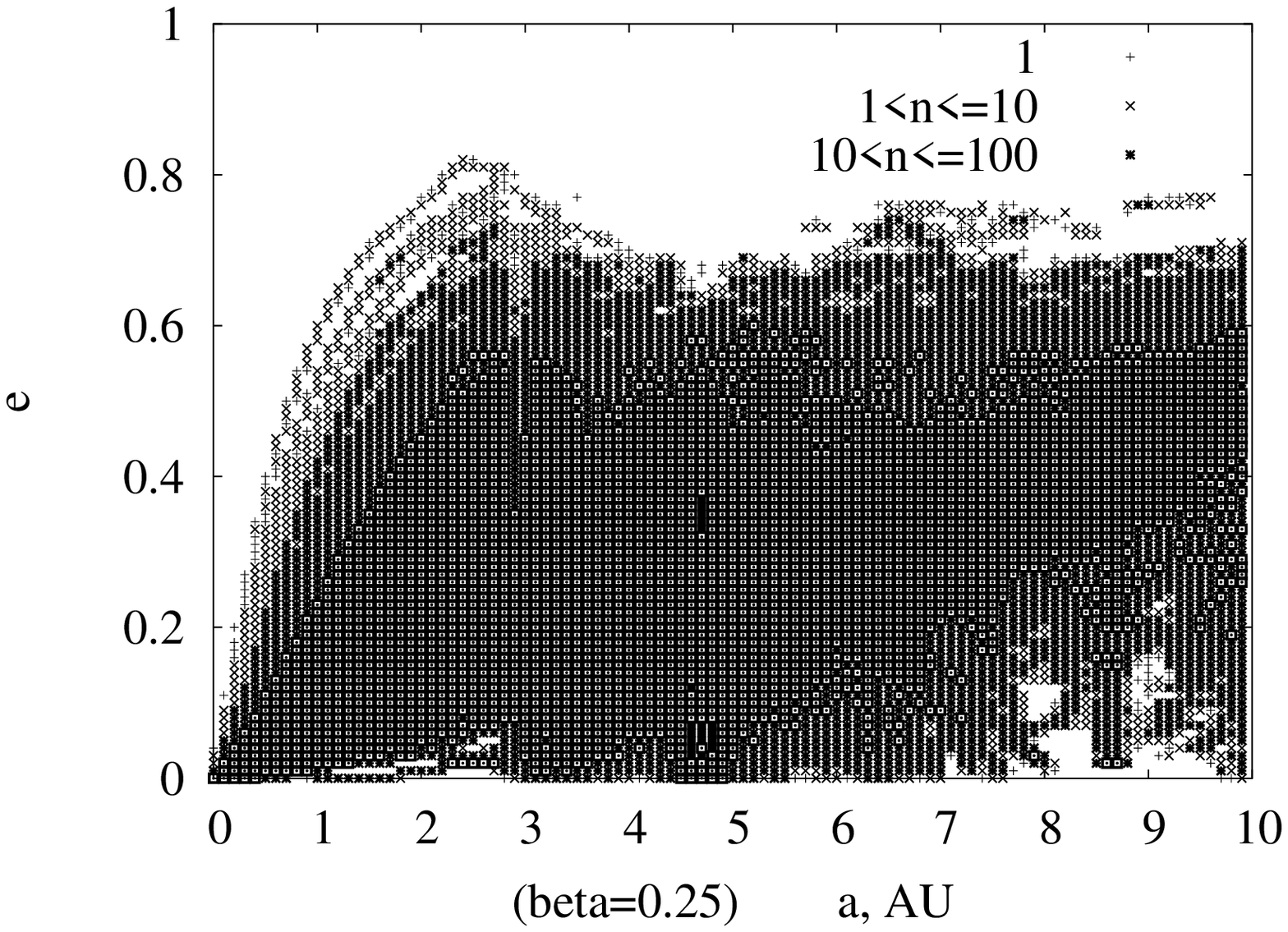}
\includegraphics[width=90mm]{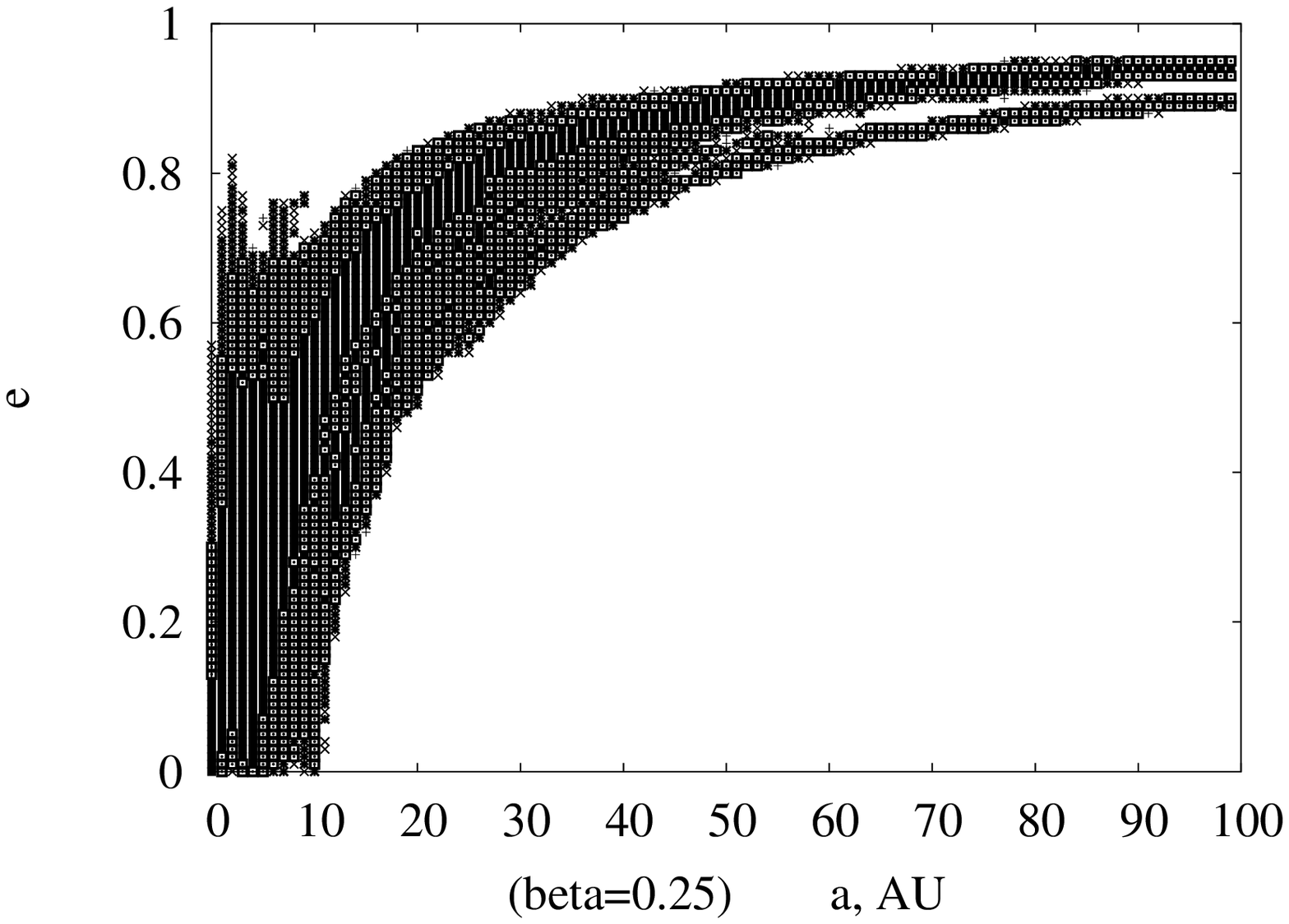}
\includegraphics[width=90mm]{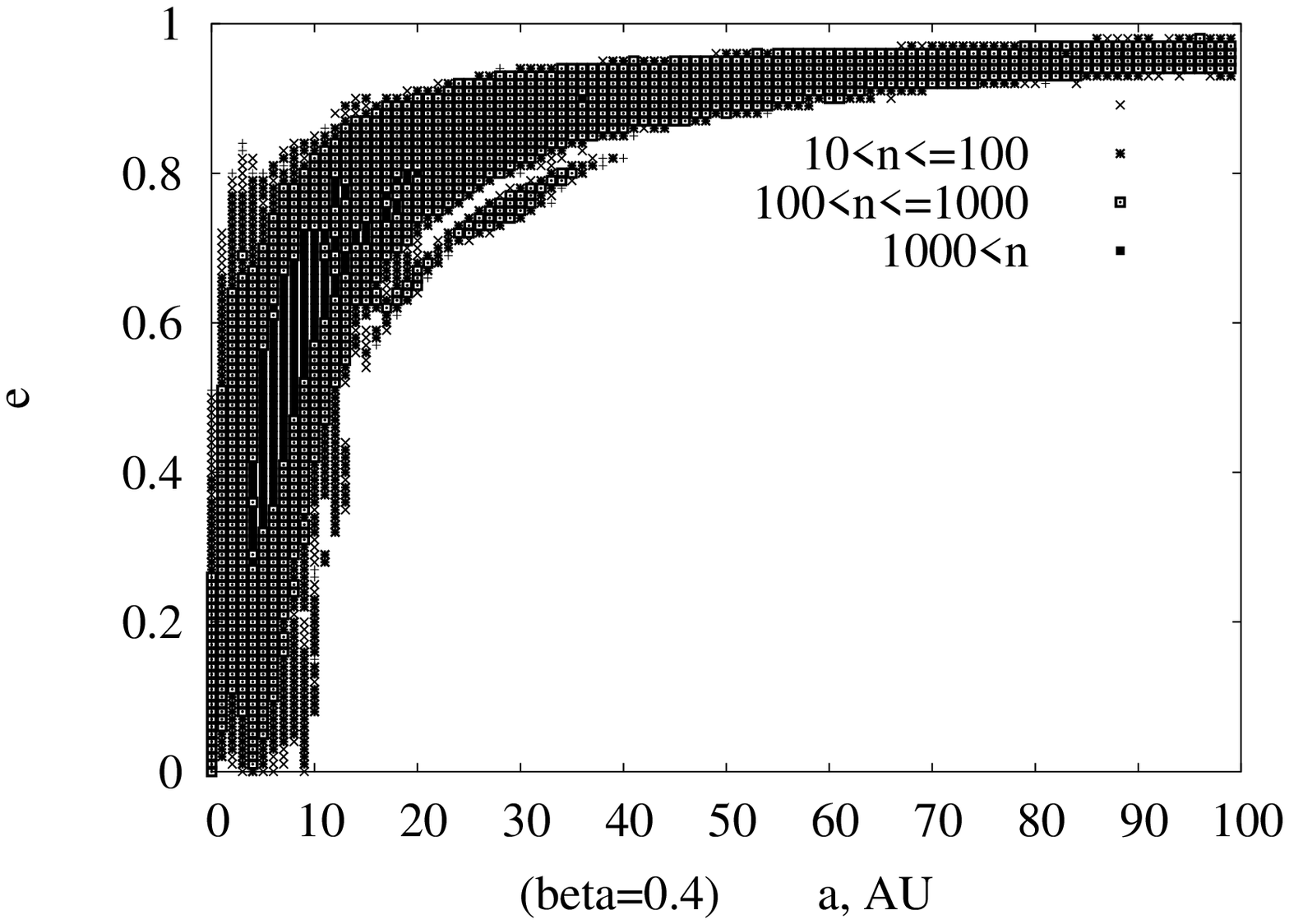}

\caption{Distribution of asteroidal dust particles with semi-major axis and eccentricity
(designations of the number of particles in one bin are 
the same in Figs. 3-7).
}

\end{figure}%

\begin{figure}

\includegraphics[width=90mm]{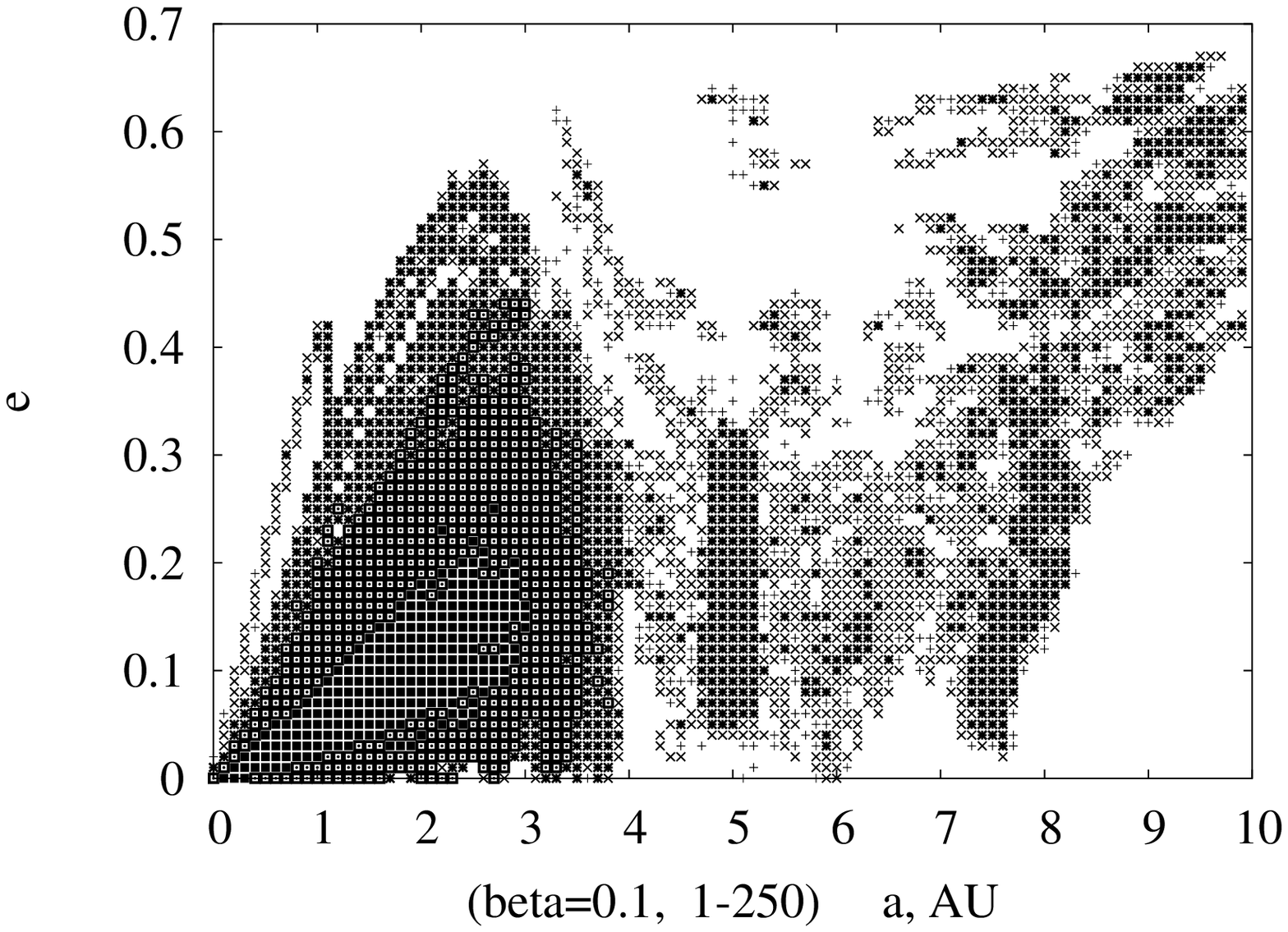}
\includegraphics[width=90mm]{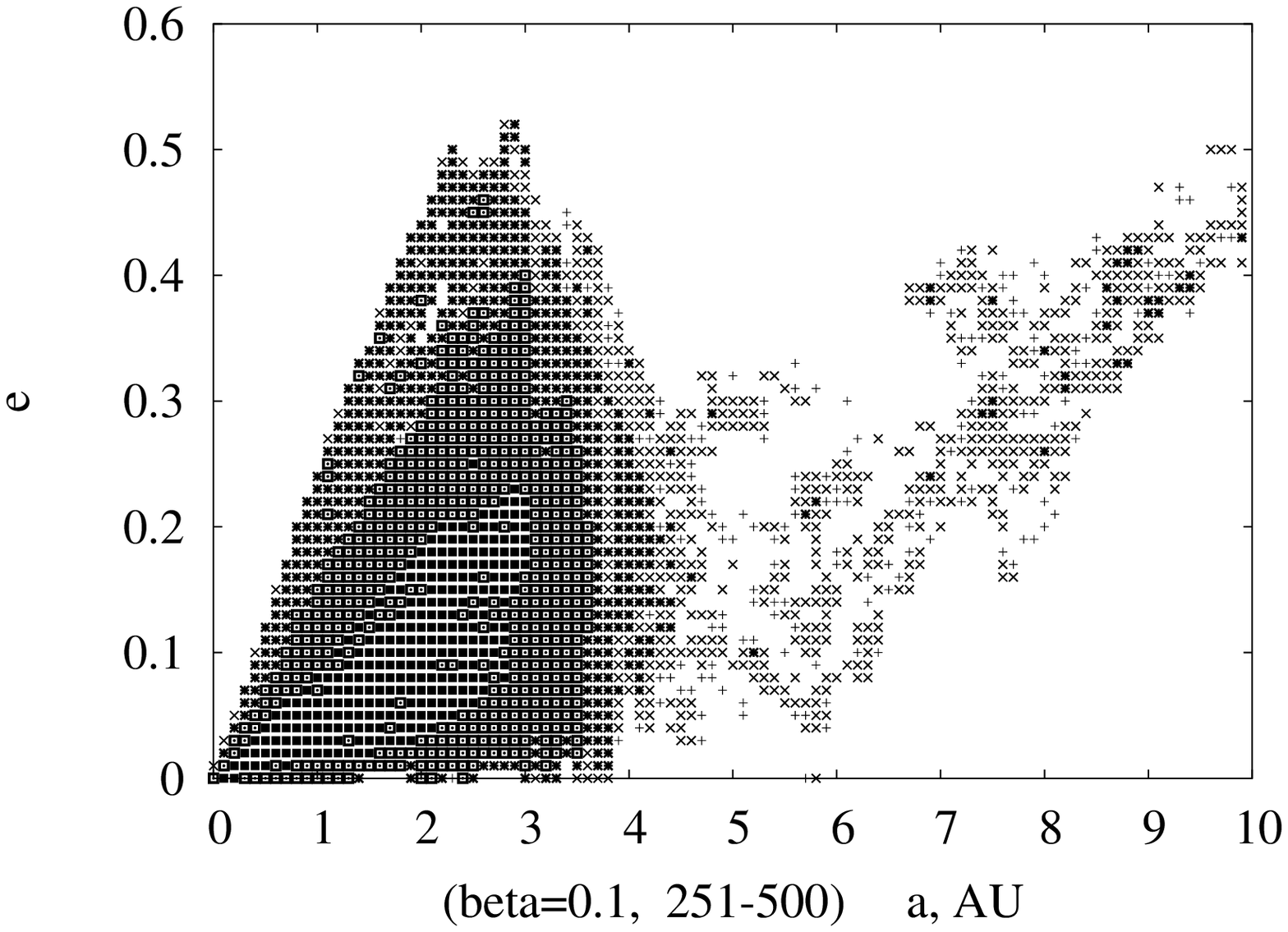}
\includegraphics[width=90mm]{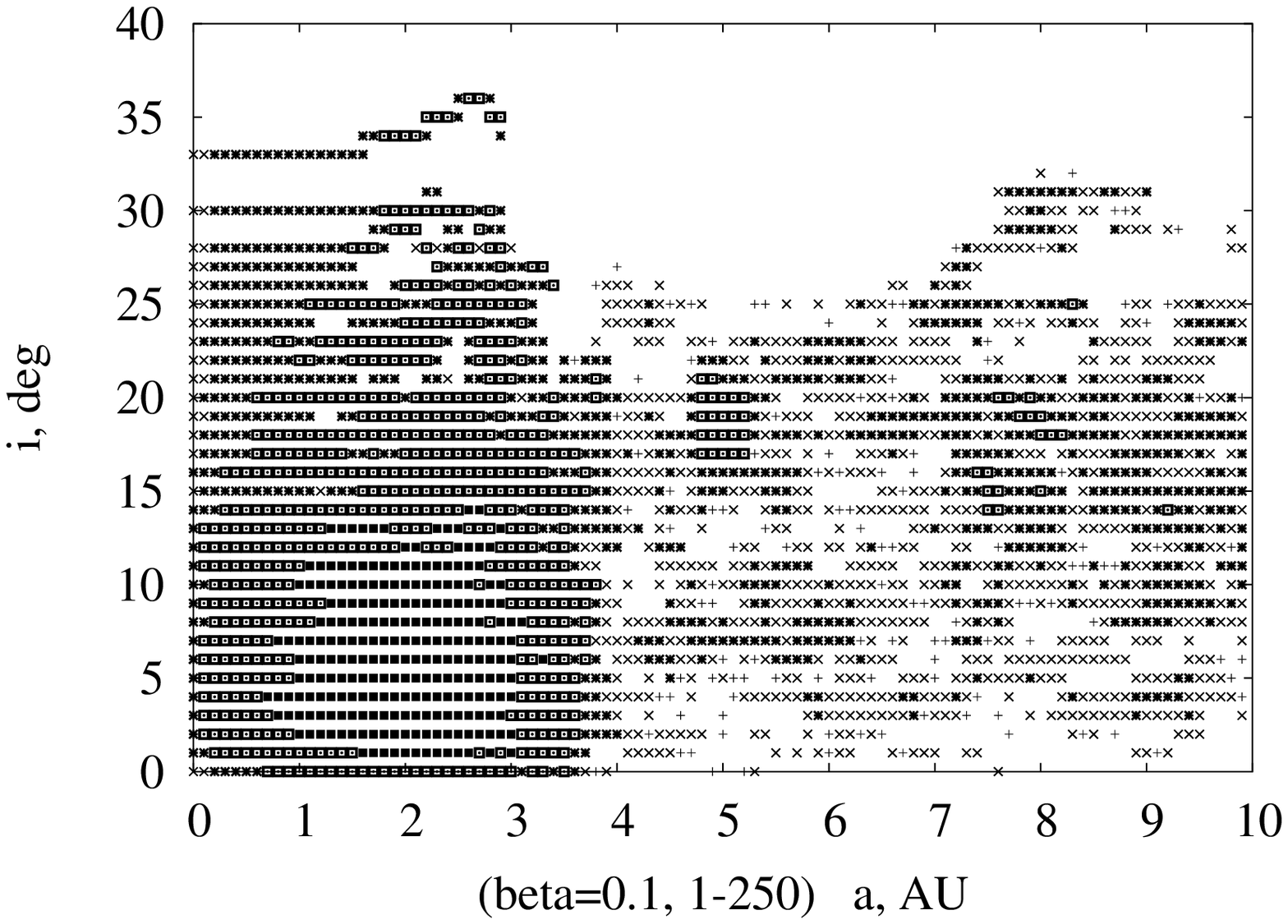}
\includegraphics[width=90mm]{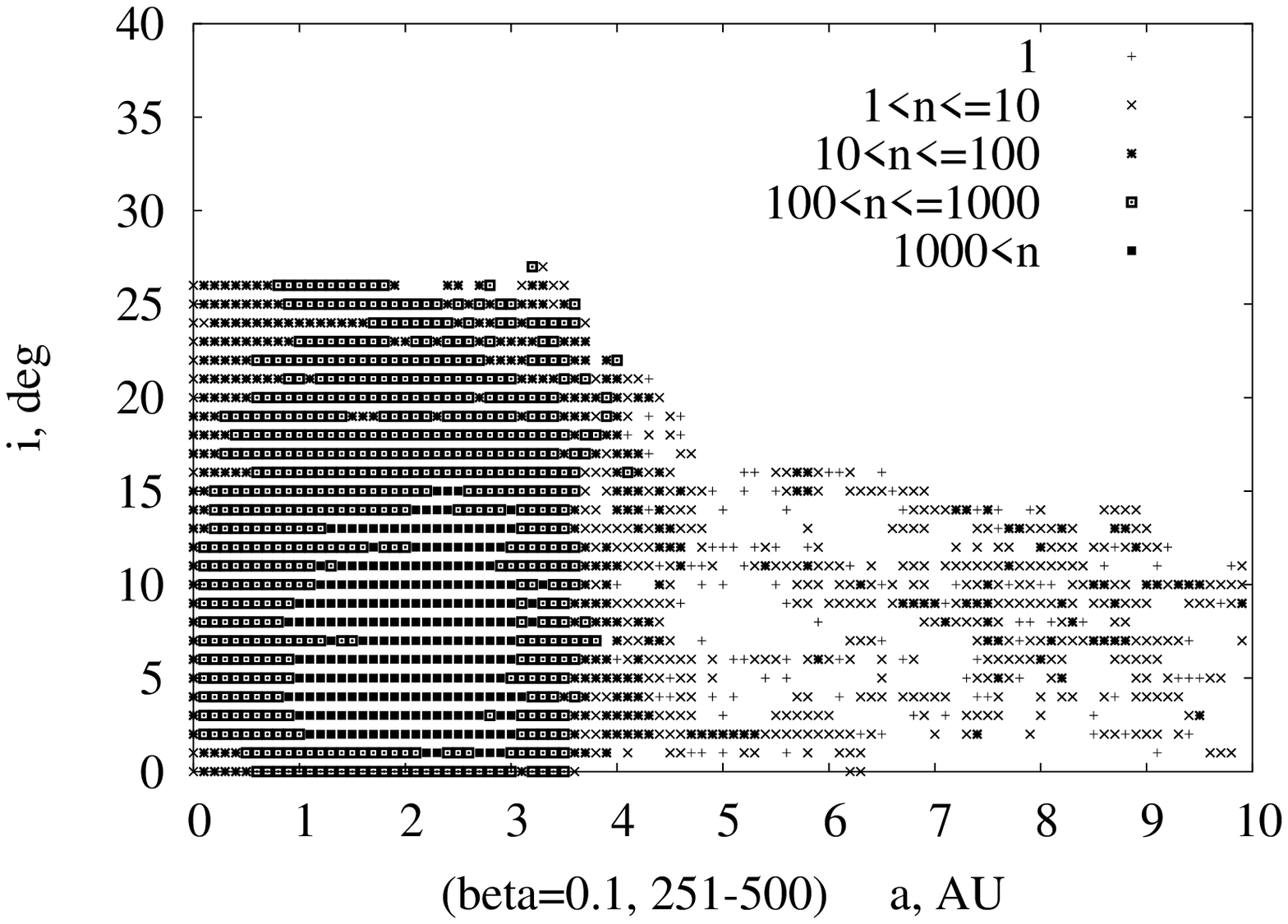}

\caption{Distribution of asteroidal dust particles with semi-major axis and eccentricity,
and with semi-major axis and inclination for $\beta$=0.1. The left figures correspond 
to the runs with initial positions and velocities close to those of the first 250 numbered
main-belt asteroids, and the right figures correspond to the runs 
with initial positions and velocities 
close to those of the asteroids with numbers 251-500 (JDT 2452500.5). 
}

\end{figure}%

\begin{figure}

\includegraphics[width=60mm]{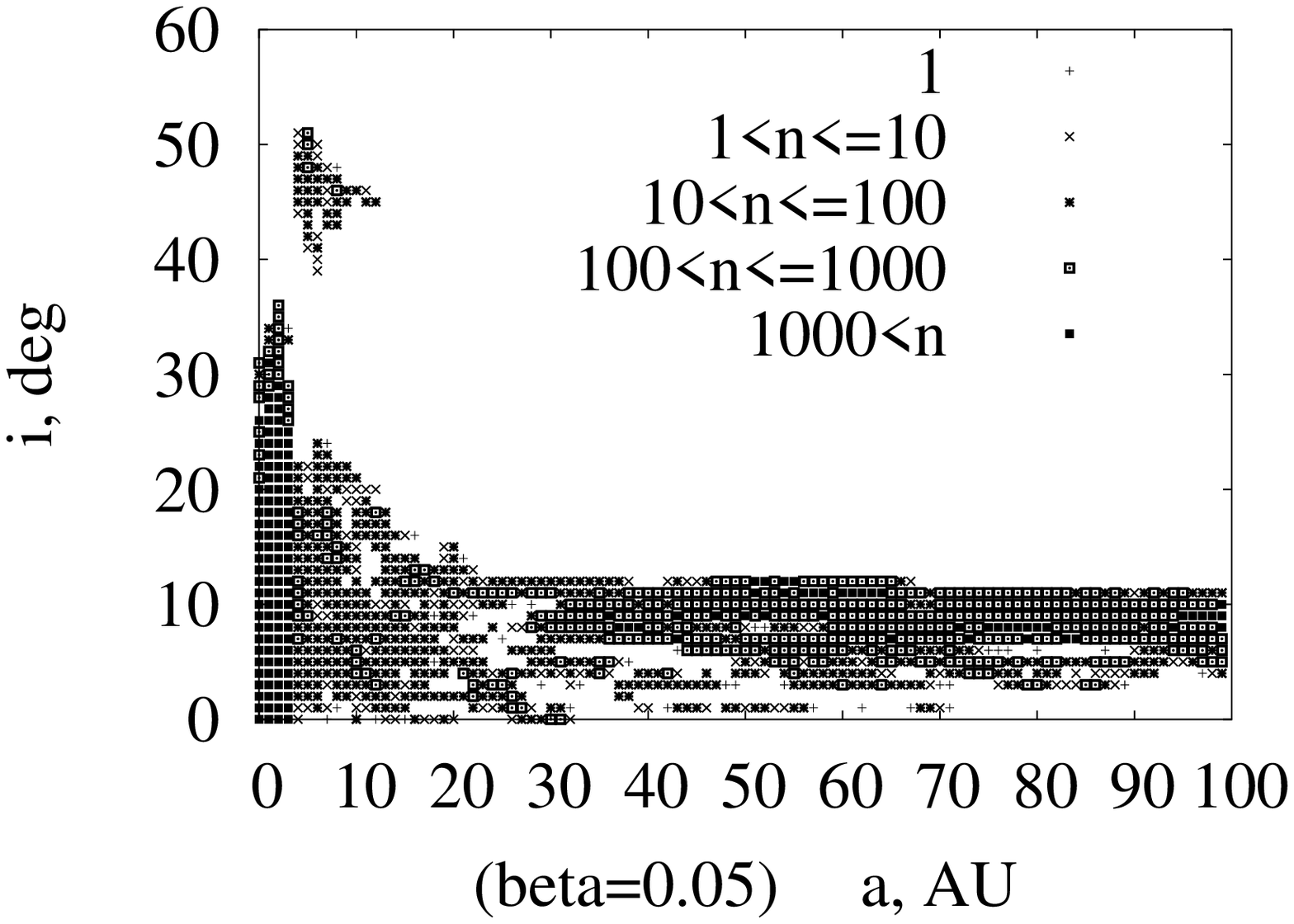}
\includegraphics[width=60mm]{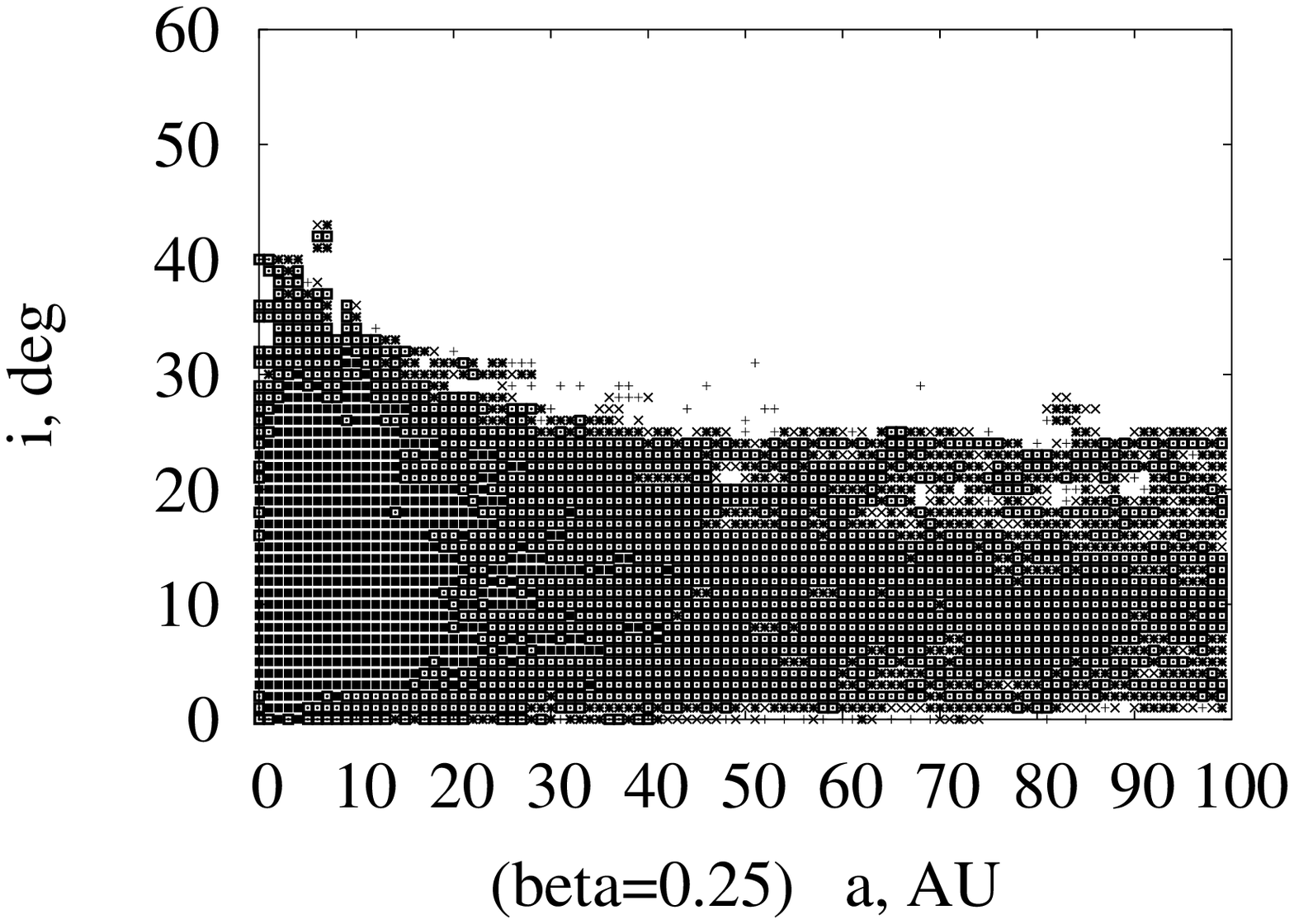}
\includegraphics[width=60mm]{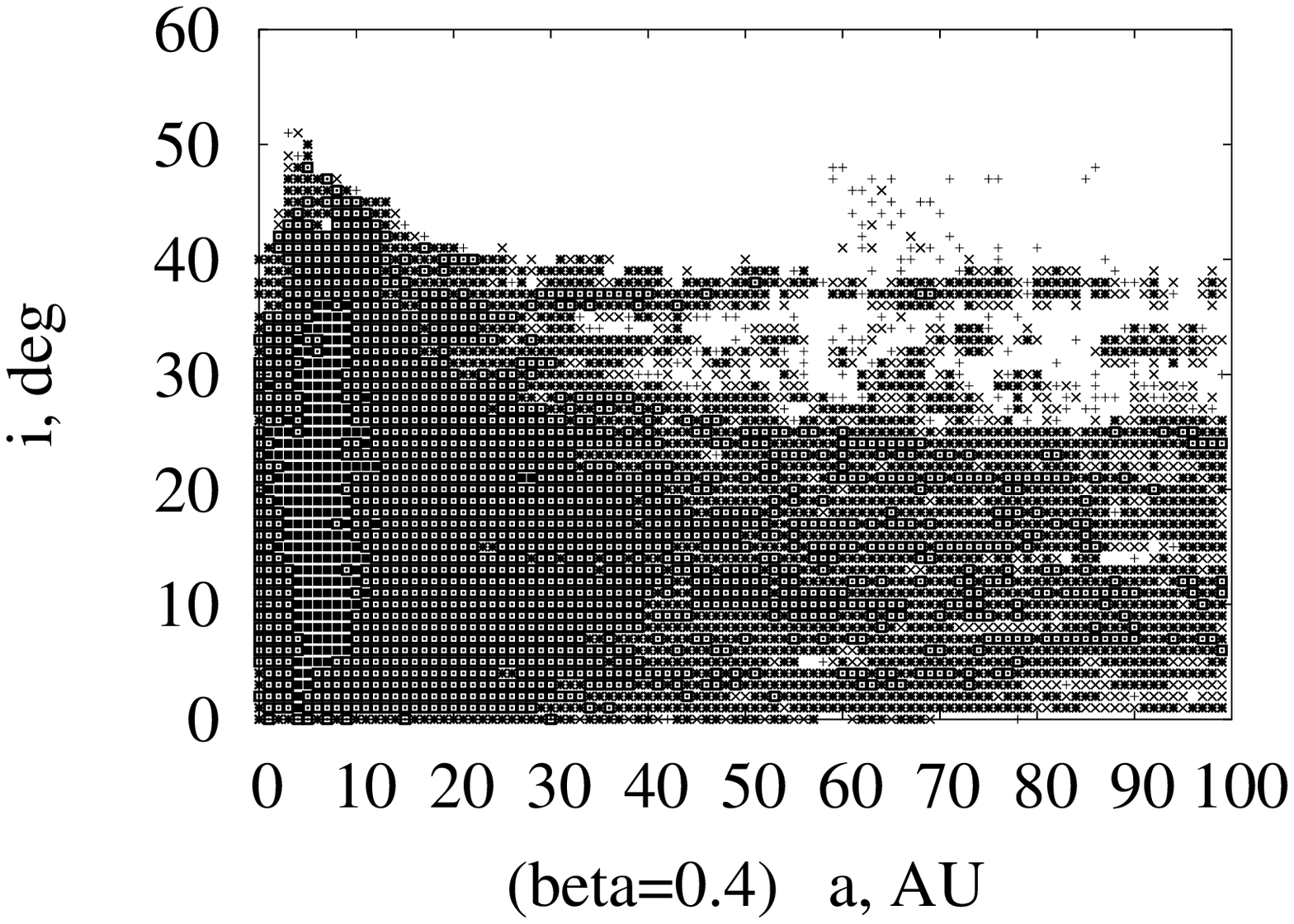}
\caption{Distribution of asteroidal dust particles with semi-major axis and inclination.
}

\end{figure}%

\begin{figure}
\includegraphics[width=90mm]{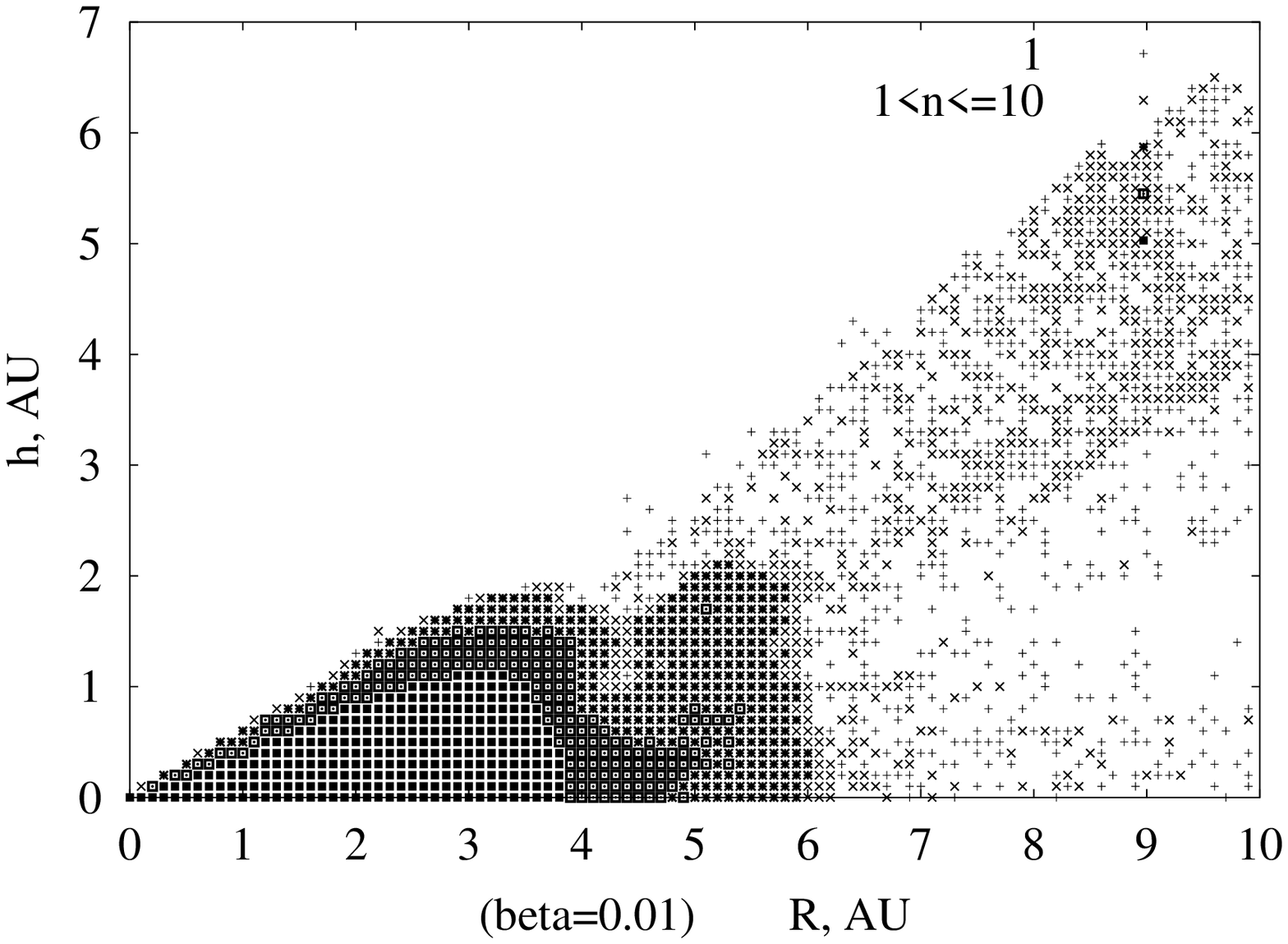}
\includegraphics[width=90mm]{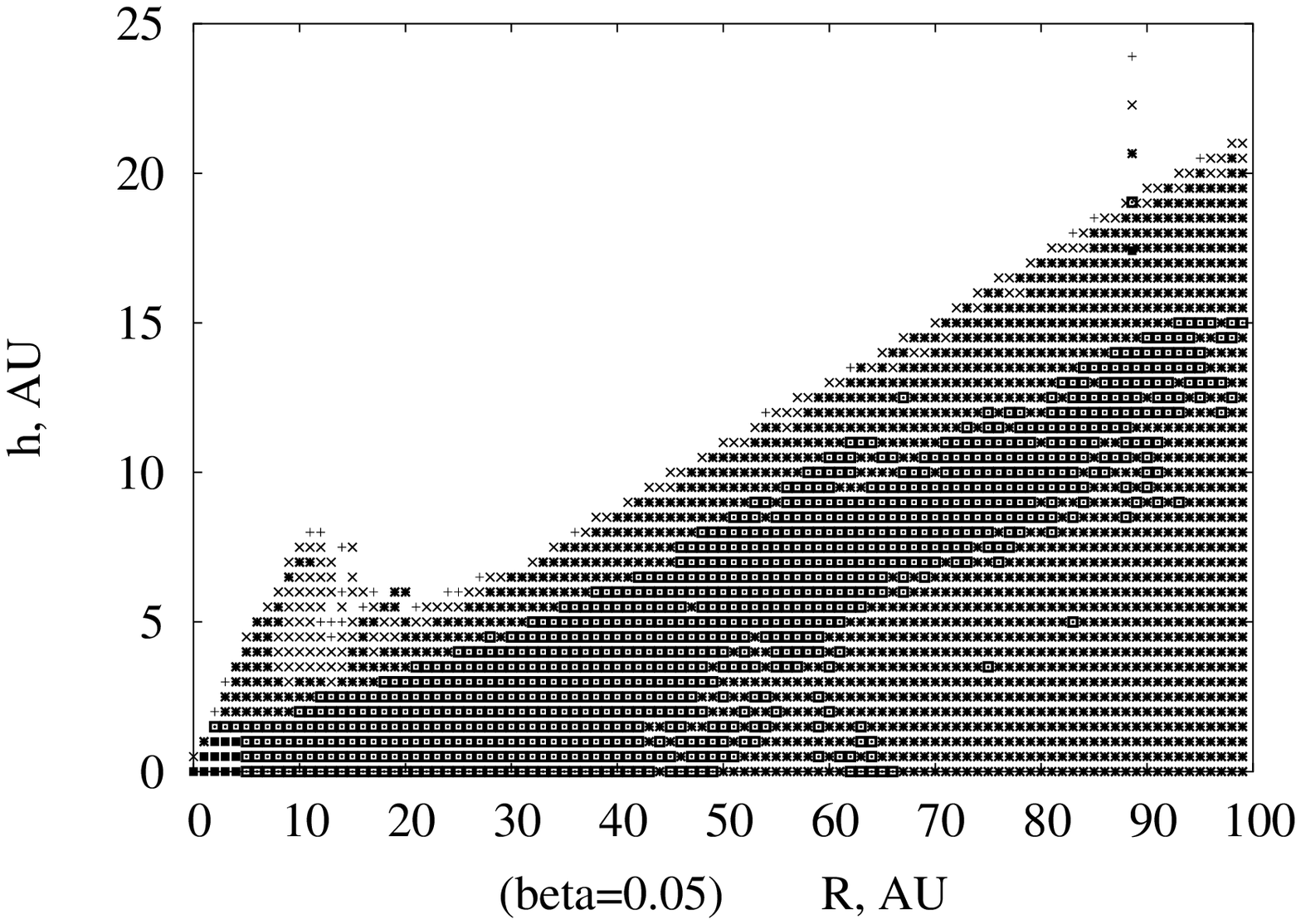}
\includegraphics[width=90mm]{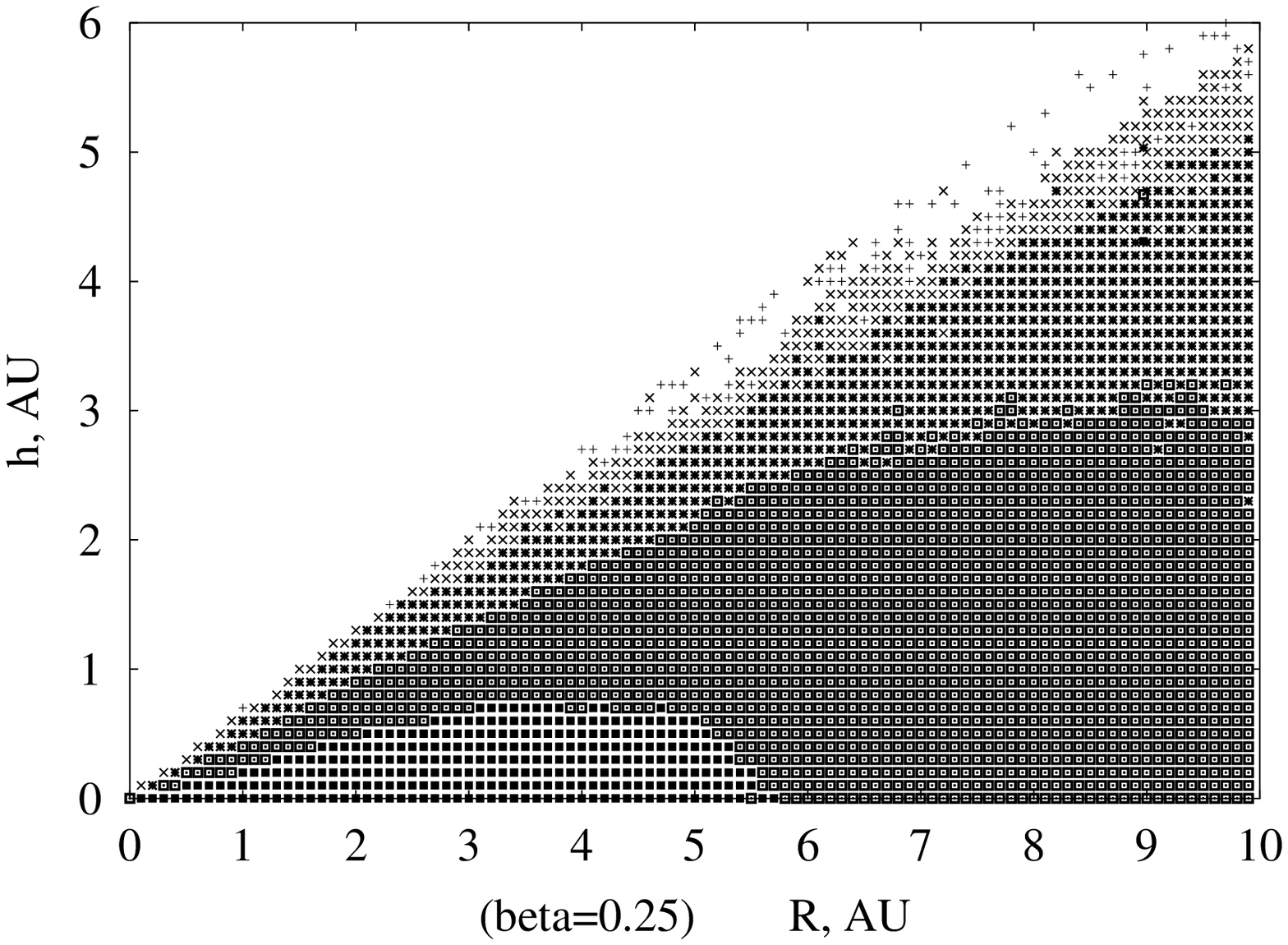}
\includegraphics[width=90mm]{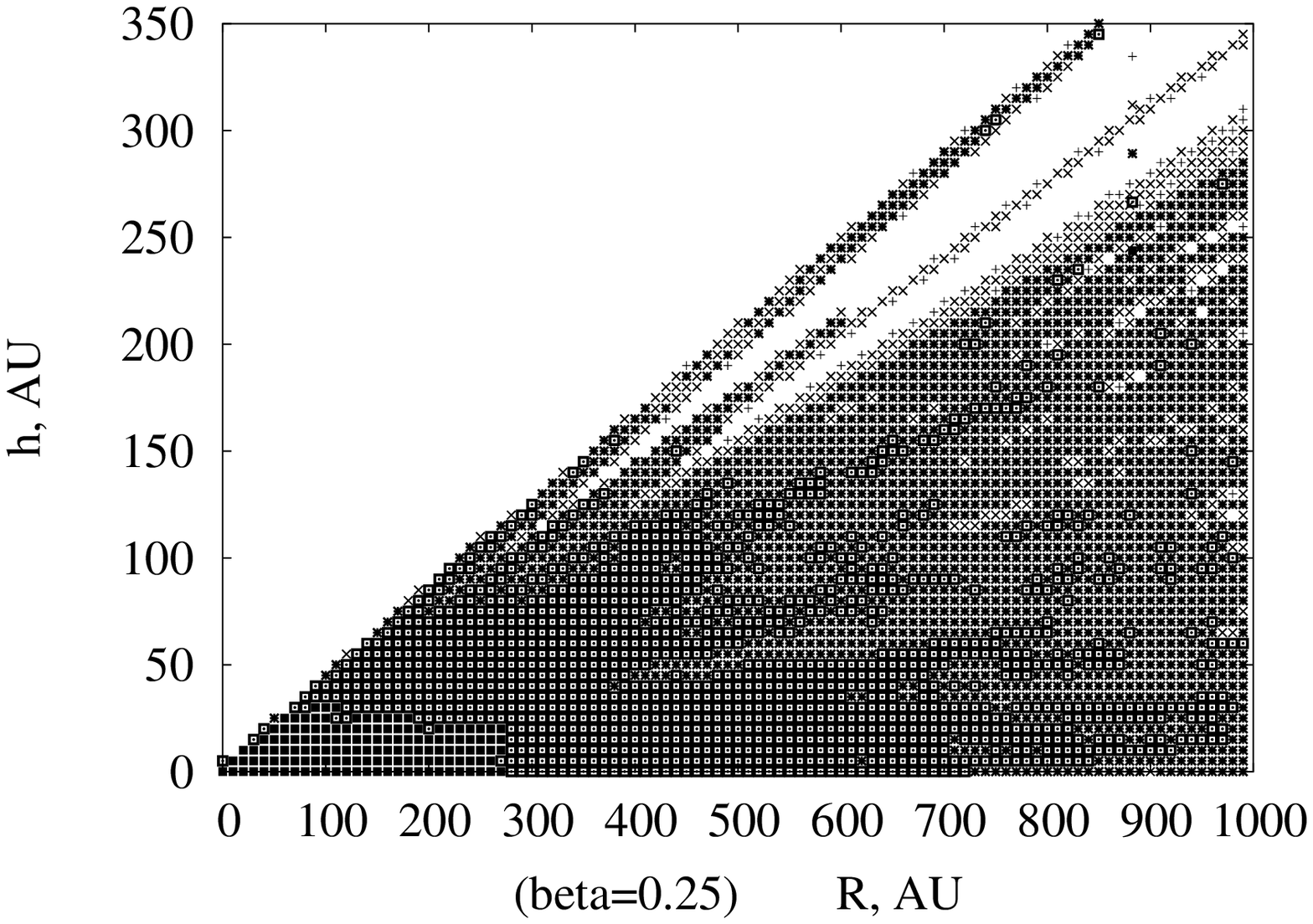}
\includegraphics[width=90mm]{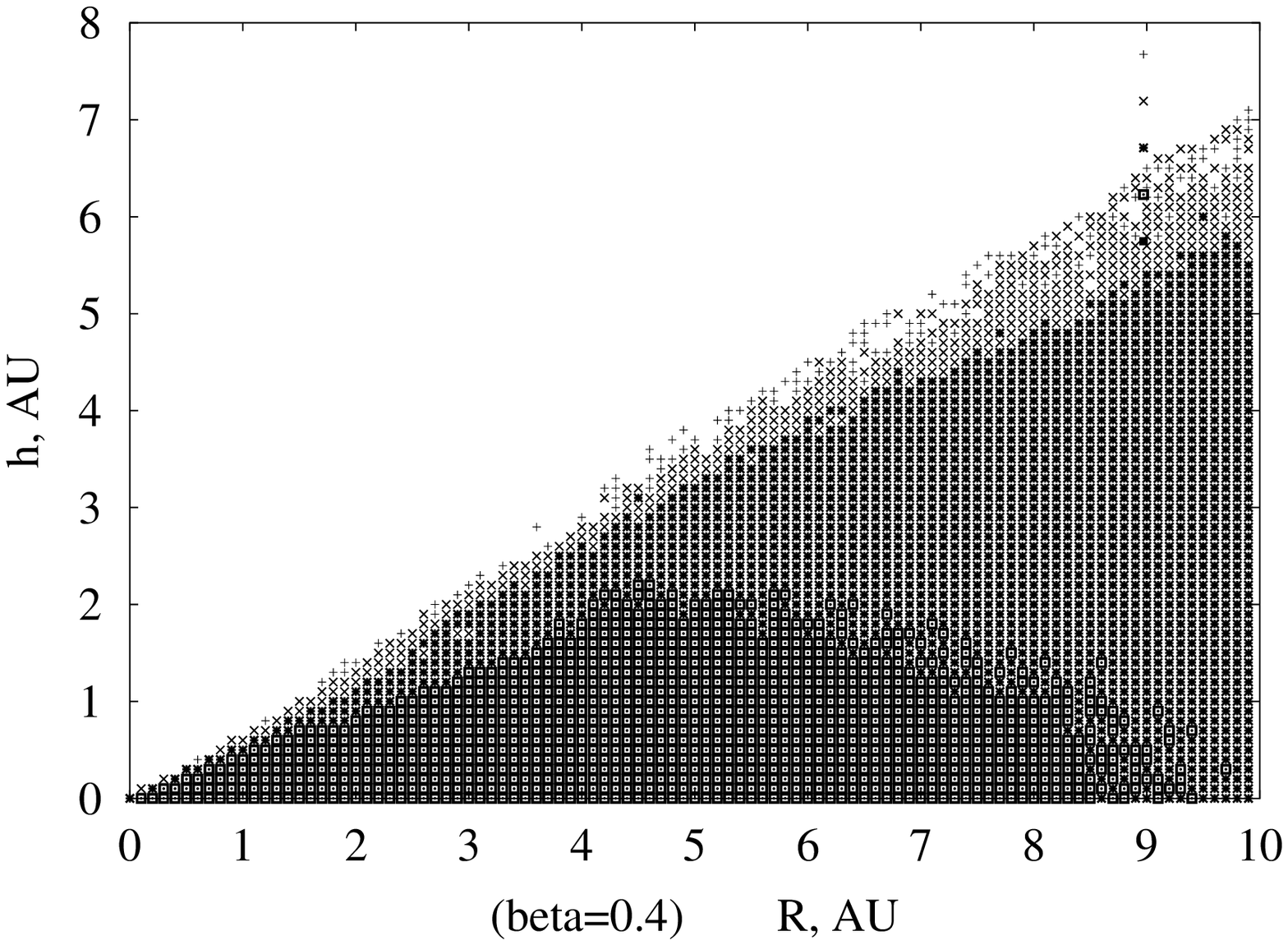}
\includegraphics[width=90mm]{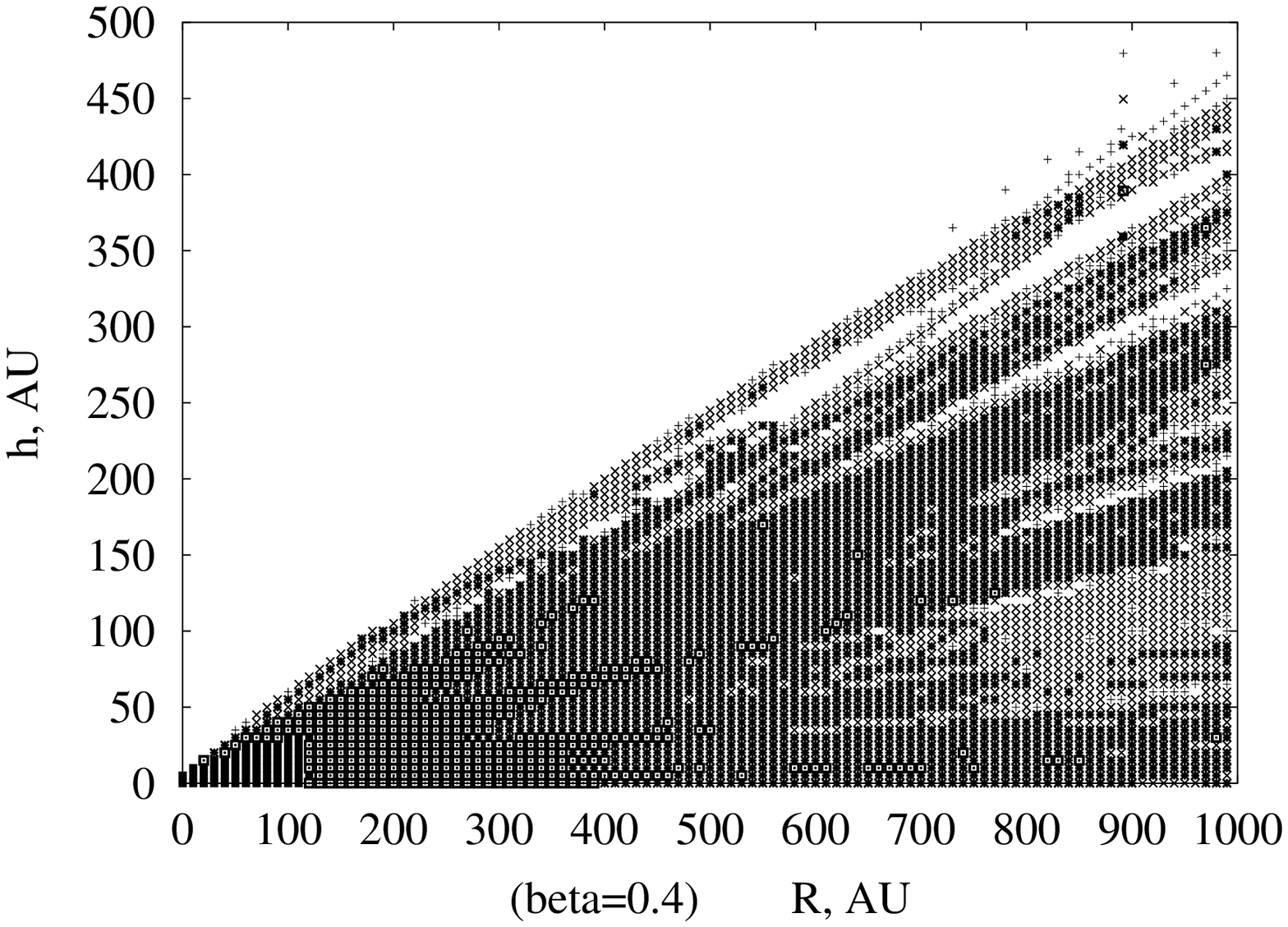}

\caption{Distribution of asteroidal dust particles with distance $R$ from the Sun and 
height $h$ above the initial plane of the Earth's orbit
(designations of the number of particles in one bin are 
the same in Figs. 3-7).
}

\end{figure}%

\begin{figure}
\includegraphics[width=91mm]{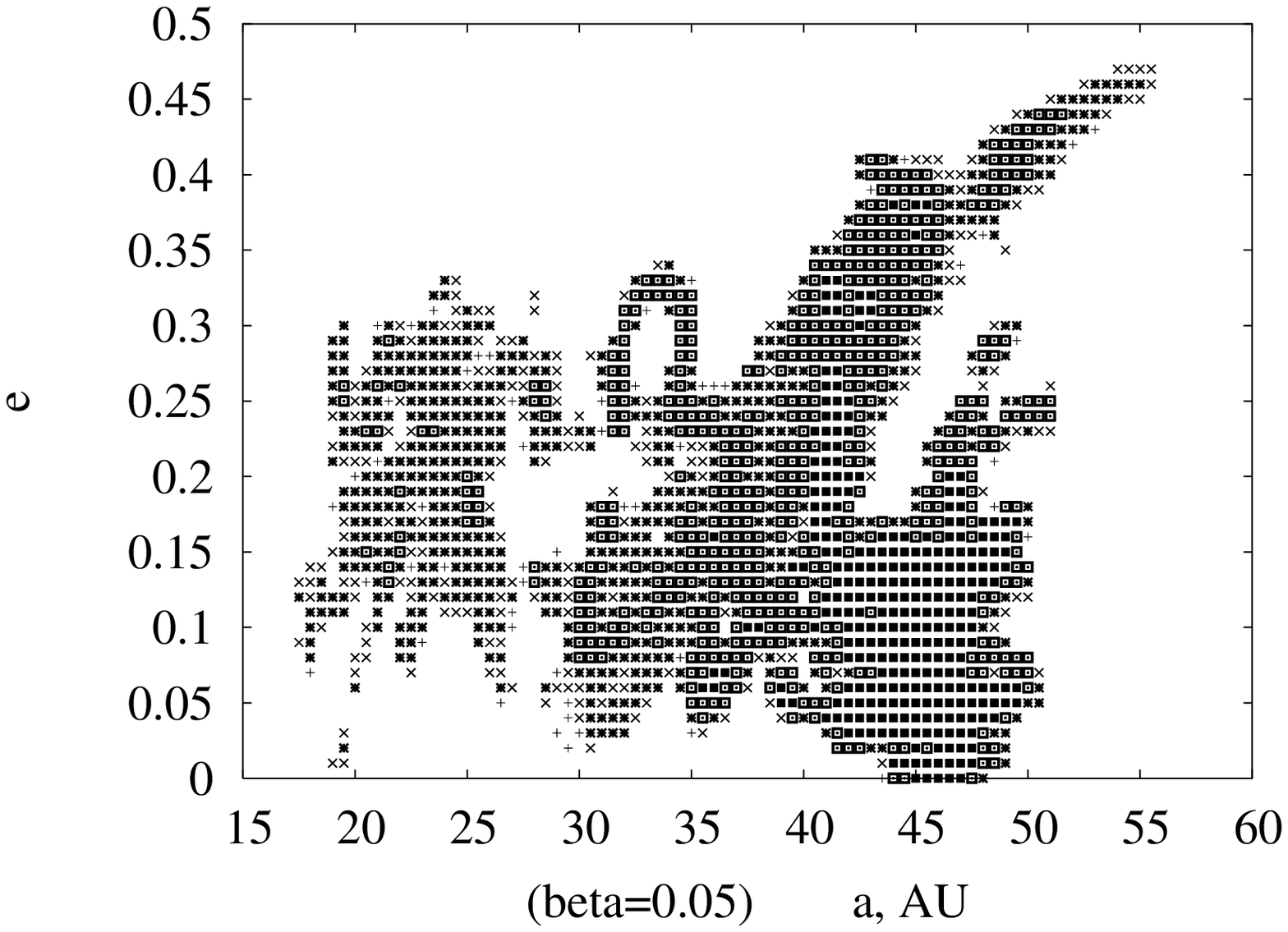}
\includegraphics[width=91mm]{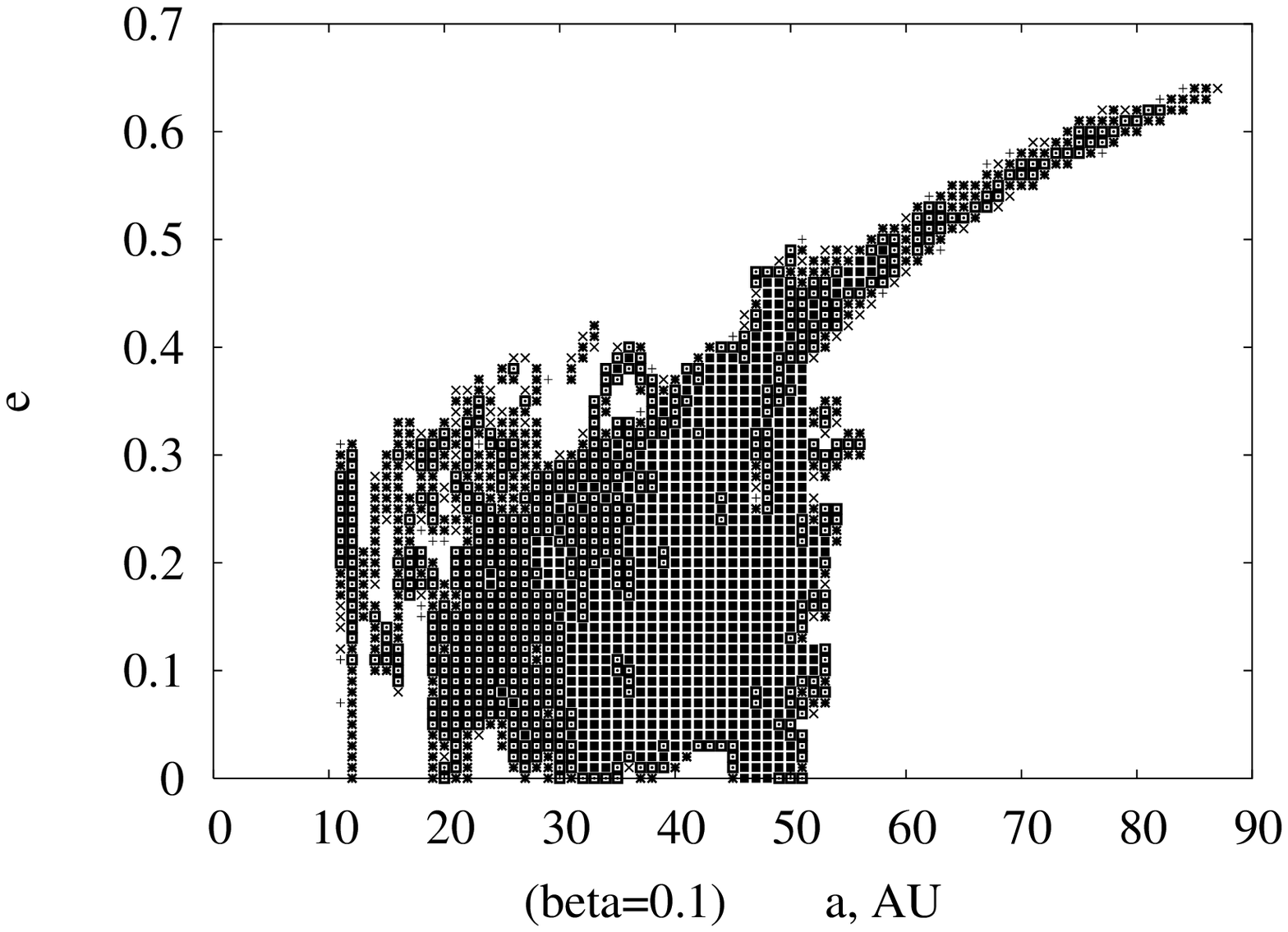}
\includegraphics[width=91mm]{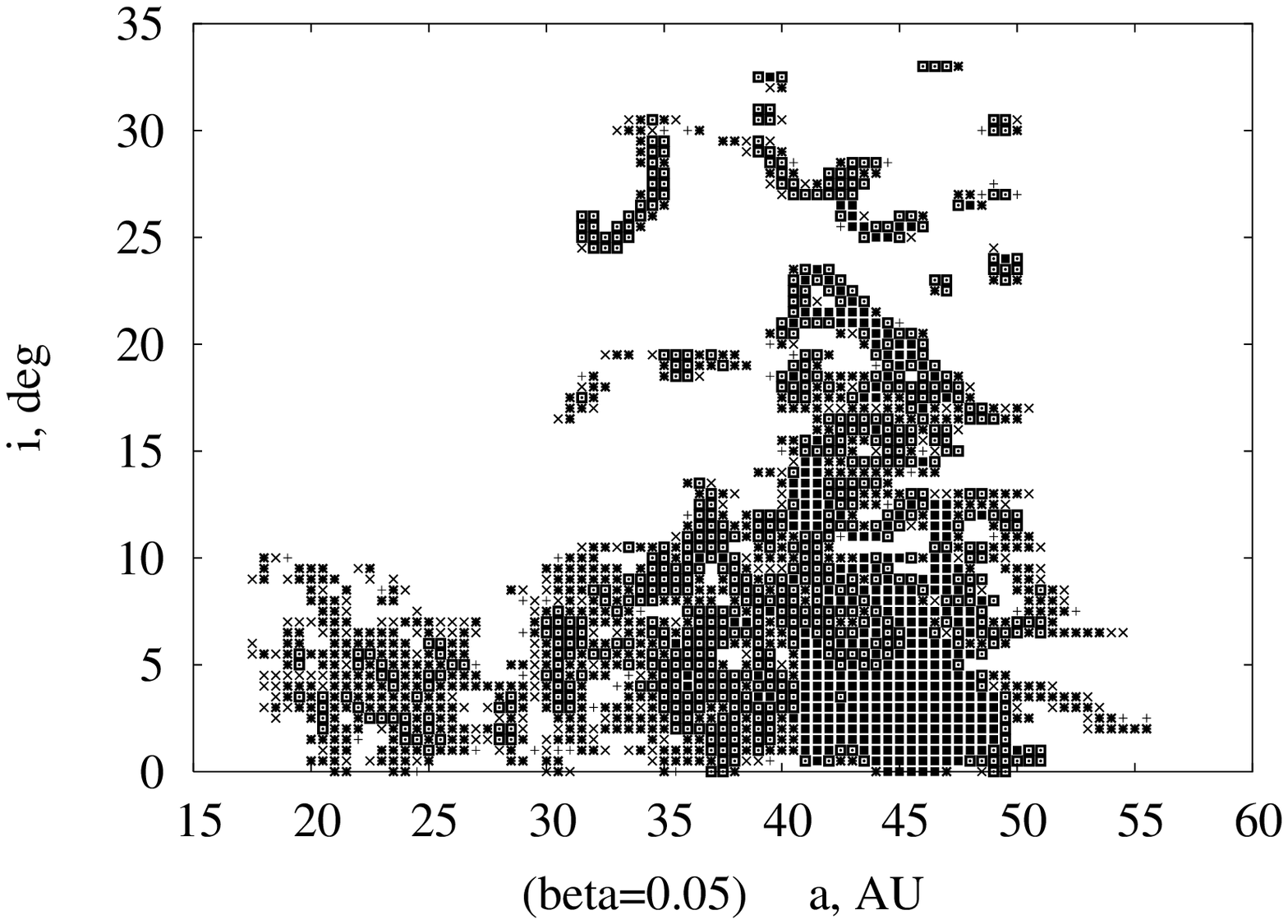}
\includegraphics[width=91mm]{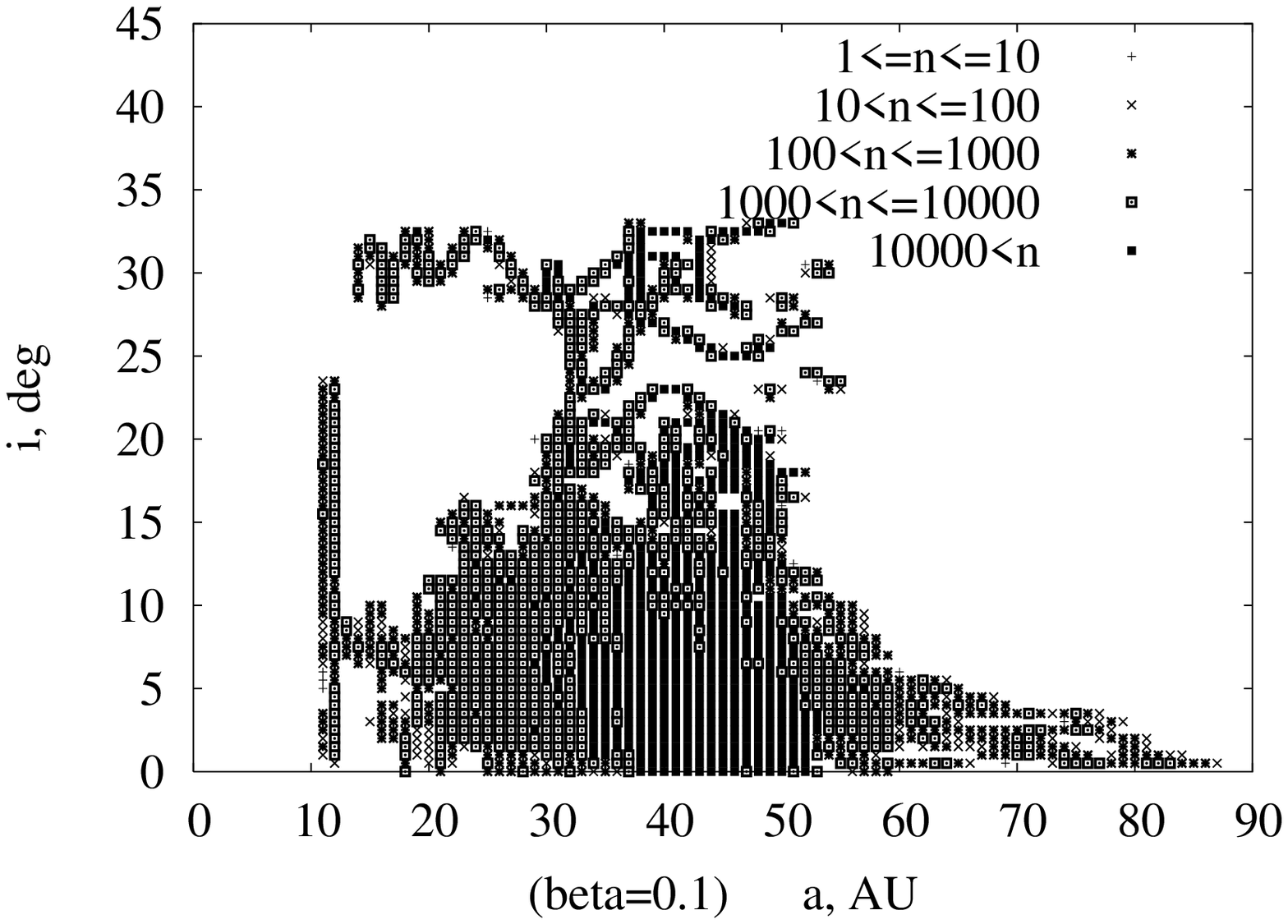} 
\includegraphics[width=91mm]{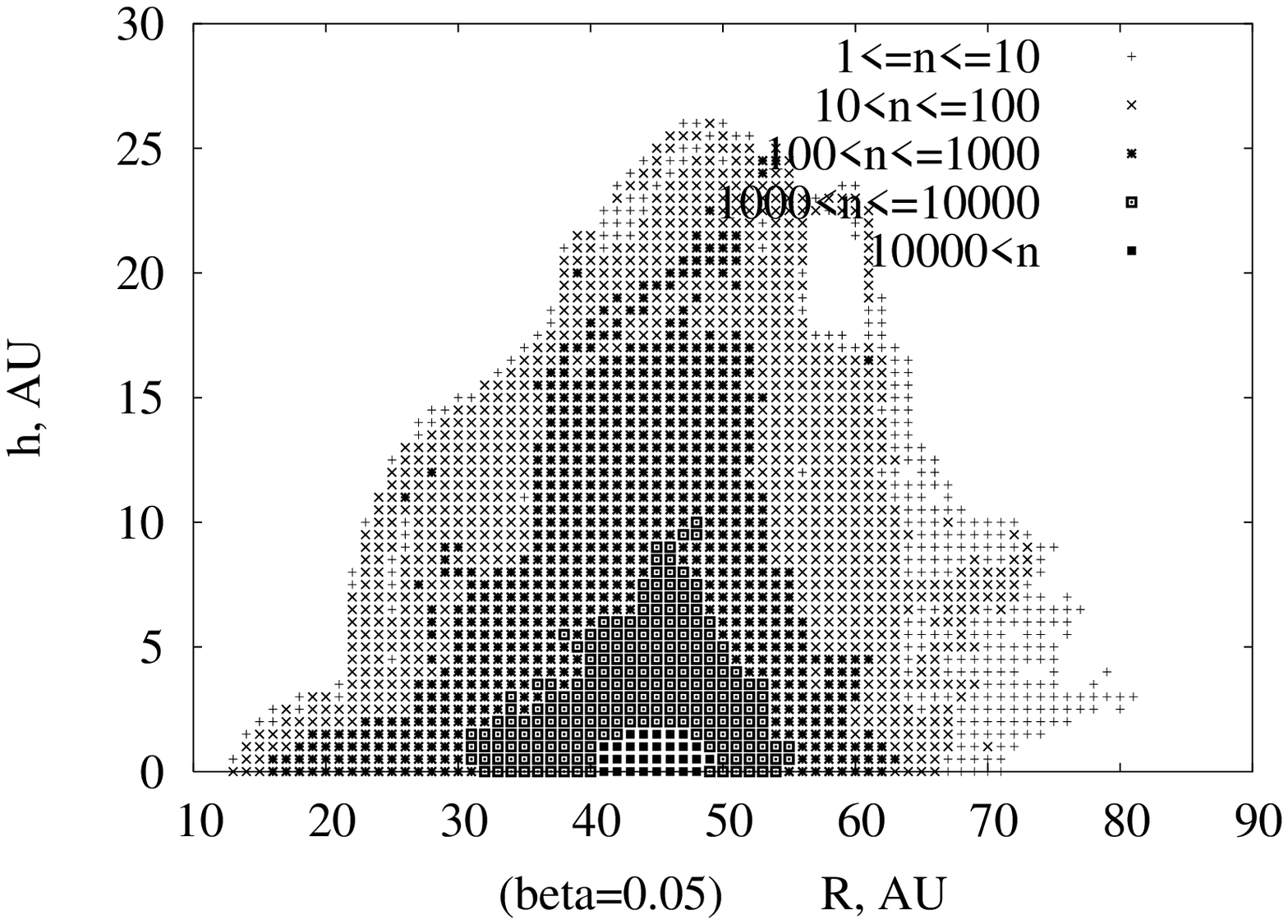}
\includegraphics[width=91mm]{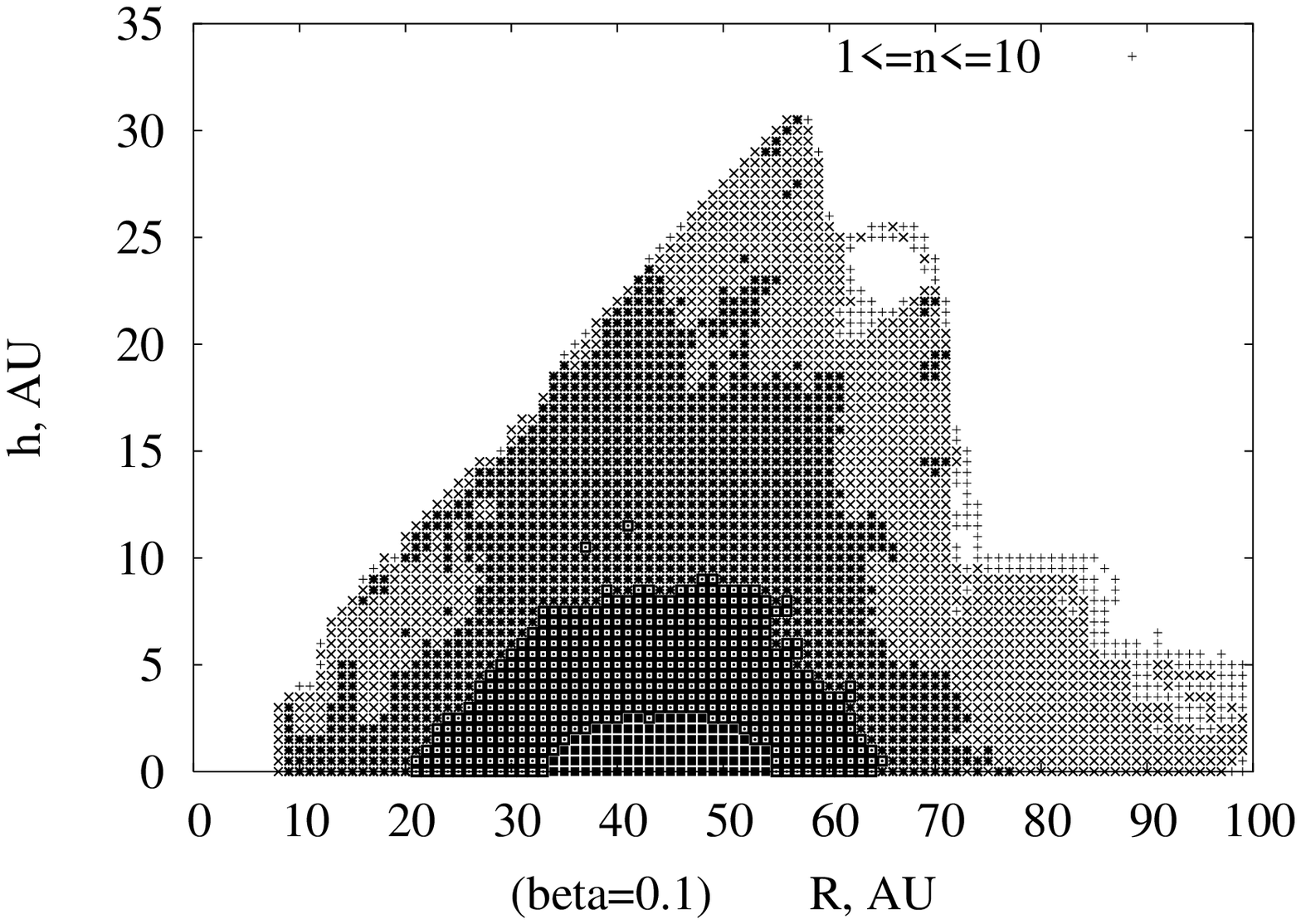}
\caption{Distribution of trans-Neptunian (kuiperoidal) dust particles with semi-major axis $a$ and 
eccenticity $e$ or inclination $i$, and 
with distance $R$ from the Sun and 
height $h$ above the initial plane of the Earth's orbit. 
The considered time interval is 2 Myr and 3.5 Myr for $\beta$=0.05 ($N$=50) and
$\beta$=0.1 ($N$=100), respectively.
}
\end{figure}%

\end{document}